\documentclass[useAMS,usenatbib]{mn2e}
\usepackage{graphicx,multirow,url}
\def\n7{NGC\,7213}
\def\fa{Fe K$\alpha$~}
\def\fb{Fe K$\beta$~}

\def\apj{ApJ}%
\def\apjl{ApJ}%
\def\apjs{ApJS}%
\def\aap{A\&A}%
\def\aaps{A\&AS}%
\def\mnras{MNRAS}%
\def\gca{Geochim.~Cosmochim.~Acta}%
\title[X-ray spectral analysis of \n7: \textit{XMM-Newton} observations]{X-ray spectral analysis of the low-luminosity active galactic nucleus \n7 using long \textit{XMM-Newton} observations\thanks{Based on observations obtained with \textit{XMM-Newton}, an ESA science mission with instruments and contributions directly funded by ESA Member States and NASA.}}
\author[D.~Emmanoulopoulos et al.]{D.~Emmanoulopoulos,$^{1}$\thanks{E-mail: D.Emmanoulopoulos@soton.ac.uk} I.~E.~Papadakis,$^{2,3}$ F.~Nicastro,$^{3,4}$ and I.~M.~M\textsuperscript{c}Hardy$^{1}$\\
$^{1}$Physics and Astronomy, University of Southampton, SO17 1BJ Southampton, United Kingdom\\
$^{2}$Physics Department, University of Crete, PO Box 2208, 71003 Heraklion, Greece\\
$^{3}$IESL, Foundation for Research and Technology, 71110 Heraklion, Greece\\
$^{4}$INAF-Osservatorio Astronomico di Roma, Via di Frascati 33, 00040 Monte Porzio Catone, Italy}
\begin{document}
\date{Accepted -- 2012 December 12. Received -- 2012 December 12; in original form -- 2012 June 29}
\pagerange{\pageref{firstpage}--\pageref{lastpage}} \pubyear{2002}
\maketitle

\label{firstpage}
\begin{abstract}
We present the X-ray spectral results from the longest \textit{X-ray multi-mirror mission-Newton} observation, 133 ks, of the low luminosity active galactic nucleus \n7. The hardness ratio analysis of the X-ray light curves discloses a rather constant X-ray spectral shape, at least for the observed exposure time, enabling us to perform X-ray spectral studies using the total observed spectrum. Apart from a neutral \fa emission line, we also detect narrow emission lines from the ionised iron species, Fe {\sc xxv} and Fe {\sc xxvi}. Our analysis suggests that the neutral \fa originates from a Compton-thin reflector, while the gas responsible for the high ionisation lines is collisionally excited. The overall spectrum, in the 0.3--10 keV energy band, registered by the European Photon Imaging Camera, can be modelled by a power-law component (with a slope of $\Gamma\simeq1.9$) plus two thermal components at 0.36 and 8.84 keV. The low-energy thermal component is entirely consistent with the X-ray spectral data obtained by the Reflection Grating Spectrometer between 0.35--1.8 keV.   
\end{abstract}


\section{Introduction}
\label{sect:intro}
\n7 (z=0.005839) is a face on Sa galaxy hosting an active galactic nucleus (AGN). It has been classified as a type I Seyfert by \citet{phillips79}, based on its H$_\alpha$ linewidth (full width at zero intensity of $13000$ km s$^{-1}$), but also as a low-ionisation nuclear emission line region (LINER) by \citet{filippenko84},  based on the study of a  variety of optical emission lines which were observed to have a full width at half maximum of  200 to 2000 km s$^{-1}$.\par
\n7 hosts a black hole with a mass of about $10^8\;{\rm M}_{\sun}$ \citep{woo02}. It has a bolometric luminosity of $L_{\rm bol}=1.7 \times 10^{43}$ ergs s $^{-1}$ \citep{emmanoulopoulos12}, yielding a rather low accretion rate of 0.0014 (i.e.\ 0.14 per cent) of the Eddington limit being intermediate between those usually found in type I Seyfert galaxies and LINER's. \n7 exhibits also intermediate radio properties between those of radio-loud and radio-quiet AGN. Its long-term X-ray spectral shape evolution, studied with the use of \textit{Rossi X-ray Timing Explorer} (\textit{RXTE}) data, show a clear `harder when brighter' X-ray behaviour \citep{emmanoulopoulos12}. Both these facts make \n7 an extragalactic analogue of the Galactic `hard state' sources. At the same time its radio and X-ray luminosities can be fitted with the global fundamental plane being also consistent with a `hard state' identification of this AGN \citep{bell11}.\par
\n7 has been observed several times in the X-ray band. \citet{bianchi03} used simultaneously obtained observations from the European Photon Imaging Camera (EPIC), on board \textit{X-ray multi-mirror mission-Newton} (\textit{XMM-Newton}) observatory, and \textit{BeppoSAX} satellites. They found no evidence of a Compton reflection component and detected an `iron line complex' between the energies of 6.5 and 7 keV, which could be explained in terms of three narrow emission lines, Fe K$\alpha$, Fe {\sc xxv} and Fe {\sc xxvi}, produced either in the Compton-thin torus or in the broad line region (BLR). Based on the Reflection Grating Spectrometer (RGS) data, of the same \textit{XMM-Newton} pointing (around 46 ks), \citet{starling05} did not find any absorption features (like those typically observed in type I Seyfert galaxies). Instead, they detected emission lines which could arise from a collisionally ionised thermal plasma with a temperature of about 0.18 keV. Further observations using the high-energy transmission grating (HETG), on board \textit{Chandra} observatory \citep{bianchi08}, suggested that the neutral \fa emission line can be produced in the BLR since its measured full width at half-maximum of $2400^{+1100}_{-600}$ km s$^{-1}$ was fully consistent with that of the H$_\alpha$ line ($2640^{+110}_{-90}$ km s$^{-1}$) in this object. The same authors found out that the dominant component of the Fe {\sc xxv} triplet is the resonant line suggesting an origin in collisionally ionised gas \citep{porquet00,bautista00}. A broad-band X-ray study of \n7, 0.6--150 keV using observations from \textit{Suzaku} and \textit{Swift} BAT \citep{lobban10}, derived similar results with respect to the origin of the neutral \fa line and the absence of the Compton reflection component suggesting that the inner, optically thick accretion disc, is probably absent. \citet{lobban10} also found that the dominant component of the Fe {\sc xxv} is consistent with that of forbidden transition, originating from highly ionised plasma and not from collisionally ionised gas as found by \citet{bianchi08}.\par
In this paper we report the analysis results of the longest \textit{XMM-Newton} observation, around 133 ks, of \n7 performed in November 2009. Initially, in Sect.~\ref{sect:obs_reduc} we present the data-reduction procedures for the EPIC X-ray data, consisting of the pn-charge coupled device (pn-CCD) and the two metal oxide semi-conductor (MOS)-CCDs. In the same section, we present the data-reduction details for the soft X-ray data from the RGS. Then, in Sect.~\ref{sect:lcs} we show the EPIC light curve products together with a hardness-ratio analysis. In Sect.~\ref{sect:epic_spec}, we perform the spectral analysis of the EPIC data and in Sect.~\ref{sect:rgs_spec} we investigate the spectral properties of the RGS data. A discussion together with a summary of our results can be found in Sect.~\ref{sect:discus}. The cosmological parameters used throughout this paper are: $\rm{H}_0=70$ km s$^{-1}$ Mpc$^{-1}$, $\Omega_{\Lambda}=0.73$ and $\Omega_{\rm m}=0.27$, yielding a luminosity distance to \n7 of 22.12 Mpc (for a corrected redshift, $z_{\rm corr.3K}=0.005145$ to the reference frame defined by the 3 K cosmic microwave background radiation). This value of the luminosity distance appears to be fully consistent with the redshift-independent Tully-Fisher distance \citep{tully88}. Throughout this paper the error bars in the figures correspond to the 68 per cent confidence limits of the corresponding plot points.

\section{OBSERVATIONS AND DATA-REDUCTION}
\label{sect:obs_reduc}
\subsection{EPIC-MOS and EPIC-pn data-reduction}
\label{ssect:epic_obs_reduc}
\n7 was observed by \textit{XMM-Newton} (Obs-ID: 0605800301) from 2009 November 11, 21:05:46 (UTC), to 2009 November 13, 09:54:25 (UTC) (on-time: 132519 s). The pn camera was operated in Prime Small Window mode and the two MOS cameras in Prime Partial W2 (i.e.\ Small Window) mode. Medium-thickness aluminised optical blocking filters were used for all EPIC cameras to reduce the contamination of the X-rays from infrared, visible, and ultra-violet light.\par
The EPIC raw-data are reduced with the \textit{XMM-Newton} {\sc scientific analysis system} ({\sc sas}) \citep{gabriel04} version 12.0.1. After reprocessing the pn and the two MOS data-sets with the \textit{epchain} and the \textit{emchain} {\sc sas}-tools respectively, we perform a thorough check for pile-up using the task \textit{epatplot}. Neither the pn nor the two MOS data sets appear to suffer from pile-up, therefore count-rates for the source and the background are extracted from circular regions of radii (in arcsec): 37.75, 45 (pn), 39.5, 179 (MOS\,1) and 43.5, 200 (MOS\,2). The corresponding centres of the circular regions for the source and the background are the following (in camera coordinates: [RAWX,RAWY]): pn -[27660,27023], [28204,23151] 3.3 arcmin of the source to the south east,  (MOS\,1) -[27663,27029], [15947,27089] 9.6 arcmin of the source to the west-, (MOS\,2) -[27653,27031], [18569,18387] 10.7 arcmin of the source to the south west-. Note that these apertures yield the optimum signal-to-noise ratio above 7 keV. We verify that the resulting light curves are not affected by pile-up problems.\par
\begin{figure}
\includegraphics[width=3.65in]{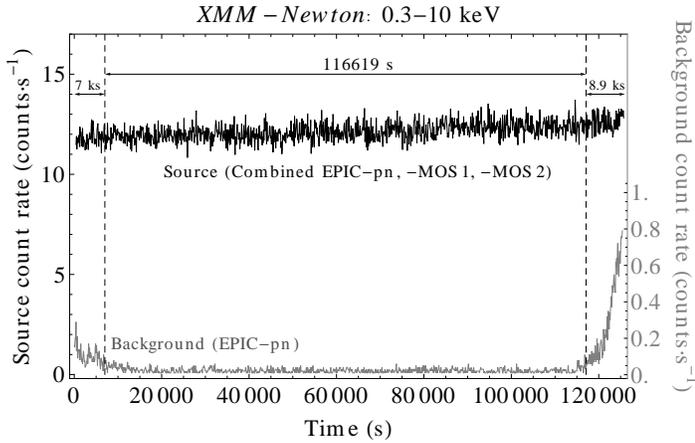}
\caption{The \n7 EPIC pn and MOS combined light curve (black-points, left-axis) and the background pn light curve (grey-points, right-axis) in the observed 0.3--10 keV energy-band. The background shows increased activity during the first 7 ks and the last 8.9 ks.}
\label{fig:lc_source_bkg}
\end{figure}

 
For the extraction of the light curves at a given energy range, we select events that are detected up to quadruple pixel-pattern on the CCDs i.e.\ PATTERN$<$12. The corrected background-subtracted light curves of the source are produced using the {\sc sas}-tool \textit{epiclccorr}. Based on the pn background light curve, in the 0.3--10 keV energy range (Fig.~\ref{fig:lc_source_bkg}, grey-points corresponding to the right-axis), an increased background activity was registered during the first 7 ks and the last 8.9 ks of the observations and thus, data from those periods are disregarded from our analysis\footnote{Despite the fact that the `high' background-states are still minimal, in terms of count-rate, with respect to the source count-rate in 0.3--10 keV (Fig.~\ref{fig:lc_source_bkg}, black-points corresponding to the left-axis), they consist mainly of events having energies more than 2 keV. At these energies the source count-rate is of the order of 1.5 counts$\cdot$s$^{-1}$.}. This leaves us with 116619 s clean exposure-time.\par
For the production of the spectra we use the {\sc sas}-task \textit{evselect} and we select for the pn camera the pulse-invariant (PI) channels from 0 to 20479, having a spectral bin size of 5 eV. For the two MOS cameras we select the PI channels from 0 to 11999 with a spectral bin size of 15 eV. For the spectral analysis we set more stringent filtering criteria from those in the case of the light curves. For the pn data we allow events that are detected up to double pixel-pattern on the CCD i.e.\ PATTERN$<$4 and we exclude all the events that are at the edge of a CCD and at the edge to a bad pixel i.e.\ FLAG$=$0. For the MOS data we allow, as in the case of the light curves, PATTERN$<$12 but we restrict to those events having also FLAG$=$0. After filtering the events, we calculate the area of the source and the background circular regions using the task \textit{backscale} and finally we compute the detectors' response matrices and the effective areas using \textit{rmfgen} and \textit{arfgen}, respectively.\par
The X-ray spectral fitting analysis has been performed by the {\sc xspec} task \citep{arnaud96} version 12.7.1, of the {\sc HEAsoft} version 6.12. In order to have similar signal-to-noise ratios among different spectral bins we group the pulse-invariant (PI) channels of the source-spectra, using the task \textit{grppha} of the {\sc ftools} \citep{blackburn95} version 6.10. The MOS source-spectra are grouped in bins of four ($\Delta E$=60 eV) and eight ($\Delta E$=120 eV) spectral channels for the energy ranges of 0.3--8 keV and 8--10 keV, respectively. The pn source-spectra are grouped in bins of four ($\Delta E=$20 eV) spectral channels for the energy range of 0.3--10 keV. All the resulting grouped spectral bins contain at least 20 photons.

\subsection{RGS Data Reduction}
\label{ssect:rgs_obs_reduc}
The RGS raw-data are processed by following the standard data-reduction threads from {\sc sas}. We use the tool \textit{rgsproc} to extract calibrated first order spectra and responses for the RGS\,1 and RGS\,2 cameras. Just like the EPIC cameras the RGS observations can be affected by high particle background periods during parts of the \textit{XMM-Newton} orbits, mostly caused by Solar activity. The high energy band is the one most affected by background flares, and high energy RGS photons are dispersed over the CCD-9 chip. Given this fact, we extract the background light curve of the CCD-9 chip and select, as good-time-intervals of the processed observations, only those during which the background count rate deviates by less than 2 standard deviations from the average background count rate of each observation. The final (i.e.\ after cleaning for high-background time intervals) exposures of the RGS\,1 and RGS\,2 spectra are 125.3 and 125.6 ks, respectively.\par 
Both RGS\,1 and RGS\,2 spectra are grouped at a resolution of 160 m\AA\ for visual purposes (i.e.\ for the figures where we present the RGS spectra). This choice corresponds to around 3.2 eV at 0.5 keV, and is equivalent to sub-sampling the effective RGS resolution by a factor of 2. Since the signal-to-noise ratio per resolution element is too low in the continuum, the chosen grouping helps to pinpoint the various emission and absorption features easier through optical inspection. For model fitting purposes, we bin the RGS spectra at a resolution of 0.015 \AA, i.e.\ we oversample the effective RGS resolution by a factor of 5.\par
Due to failures of two different read-out detector chips, early in the life of the mission, both the RGS\,1 and the RGS\,2 lack response in two different spectral intervals of 0.9--1.2 keV and 0.5--0.7 keV, respectively. We therefore consider the following two spectral intervals for spectral fitting purposes: $[(0.35-0.9) \cup (1.2-1.8)]$ keV and $[(0.35-0.5) \cup (0.7-1.8)]$ keV, for RGS\,1 and RGS\,2, respectively. The upper energy limit of 1.8 KeV is based on the fact that above this energy the RGS effective area drops from 70--100 cm$^2$ to 30 cm$^2$ and as a result the observed spectra become very noisy (carrying no extra value for model fitting purposes).\par
 The net counts per bin (i.e.\ after background subtraction) at the corresponding spectral intervals are between 100 and 300, yielding a signal to noise ratio between 10 and 17 per bin for RGS\,1 and RGS\,2, respectively. Outside this band the photon count-rate of the background spectrum becomes comparable to the background-subtracted source count-rate.\par
For the spectral fitting procedures we have used the {\em Sherpa} fitting package \citep{freeman01}, of the {\sc chandra interactive analysis of observations} ({\sc ciao}) software version 4.4, to fit simultaneously the RGS\,1 and RGS\,2 spectra of \n7.

\section{X-RAY LIGHT CURVES}
\label{sect:lcs}
\subsection{\textit{XMM-Newton} light curves}
\label{ssect:xmm_lc}
\begin{figure}
\hspace{-0.05em}\includegraphics[width=3.45in]{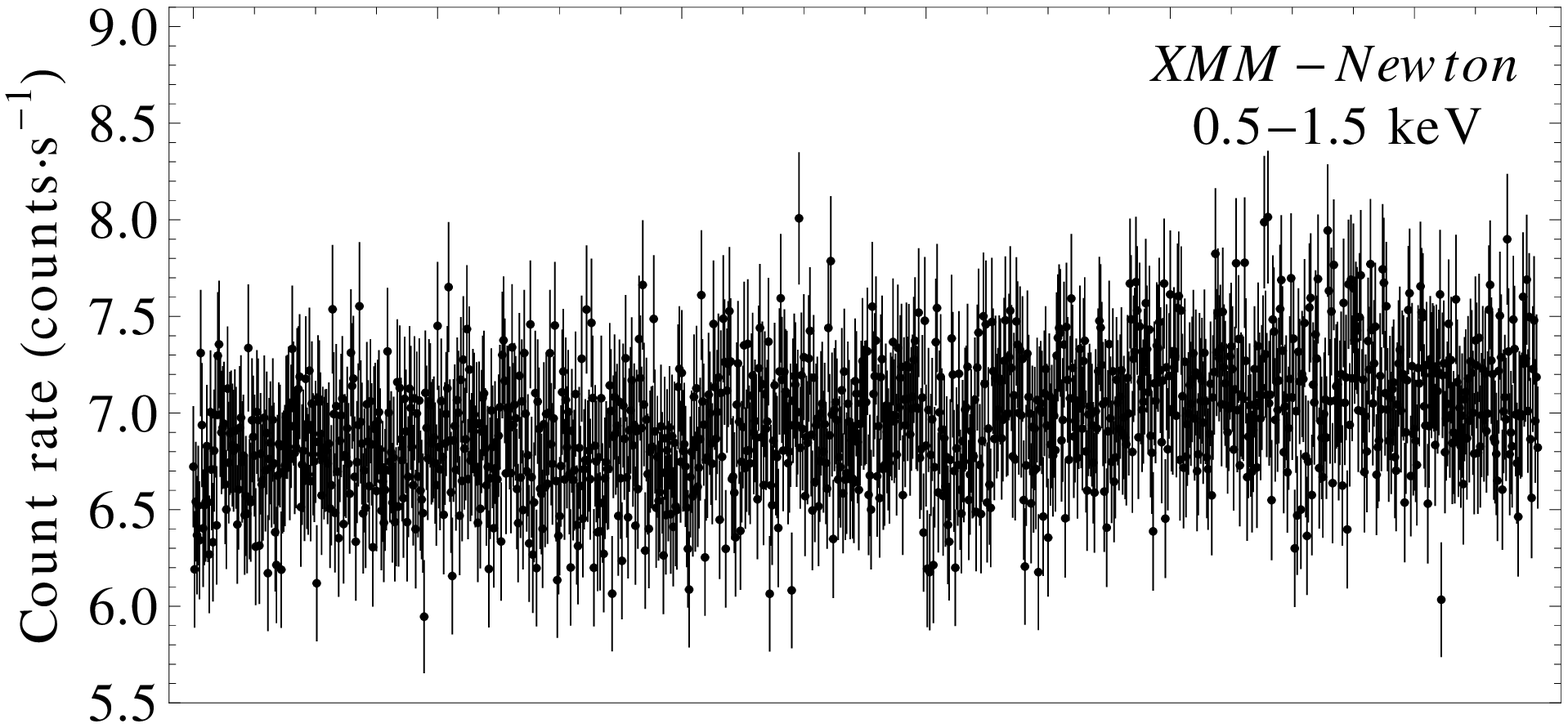}\\[-2.17em]
\hspace*{0em}\includegraphics[width=3.444in]{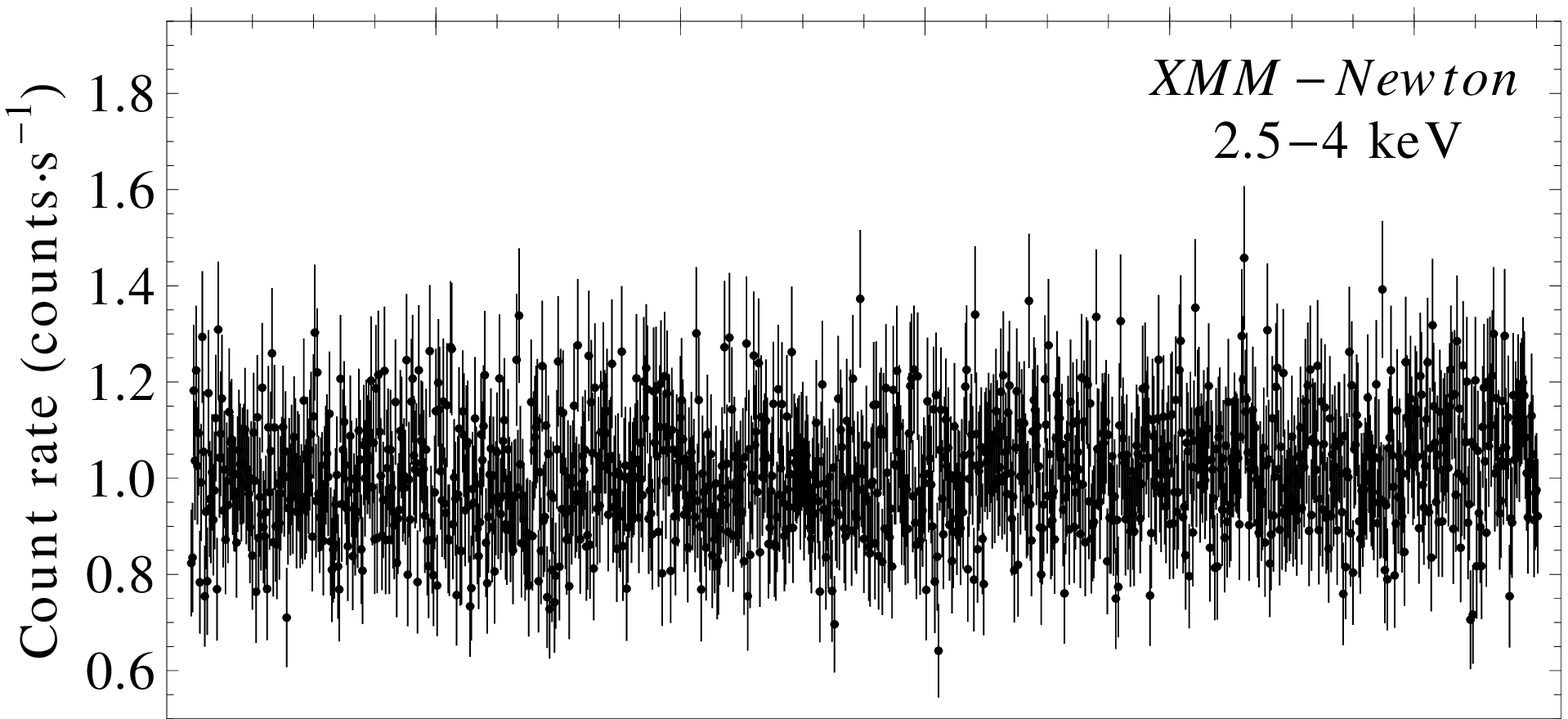}\\[-1.6em]
\hspace*{0em}\includegraphics[width=3.444in]{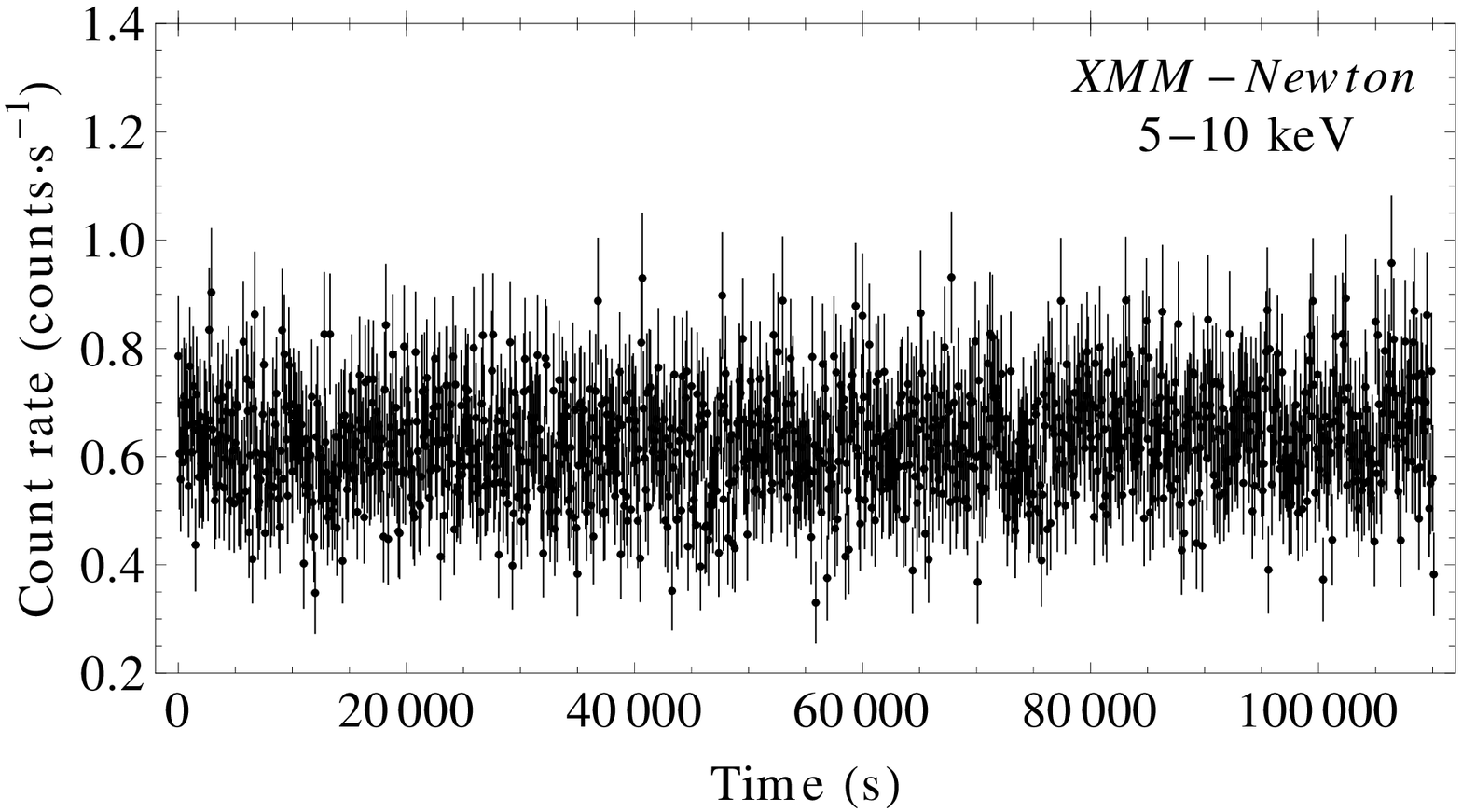}
\caption{The combined EPIC-pn,-MOS\,1,-MOS\,2 X-ray light curves of \n7 in the 0.5--1.5 keV, 2.5--4 keV, and 5--10 keV energy-bands, in bins of 100 s. Note the difference in the count-rate scale on the vertical axis.}
\label{fig:xmm_lightcurves}
\end{figure}

In order to produce the combined X-ray light curves from the EPIC-pn and the two EPIC-MOS cameras we estimate, in 100 s bins, the total count rate and the one standard deviation error bar, as the Euclidean norm of individual errors (i.e.\ the square root of the sum of the squared errors) from all three detectors, using the {\sc ftool} \textit{lcmath}. The combined final X-ray light curves in 0.5--1.5 keV, 2.5--4 keV, and 5--10 keV energy ranges, are shown in Fig.~\ref{fig:xmm_lightcurves}. A very small steady count-rate increase can be noticed in all three energy-bands, of the order of $6.7\pm0.5$, $5.7\pm1.3$ and $6.4\pm1.7$ per cent, respectively. Additionally, the fractional variability amplitudes \citep[corrected for the photon noise e.g.][]{vaughan03}, for the three bands are rather small, 1.7$\pm$0.3, 2.5$\pm$1.6 and less than 0.8 per cent\footnote{The excess variance of the light curve in the 5--10 keV energy band is negative and thus we quote the 90 per cent upper limit.}, respectively.

\subsection{\textit{XMM-Newton} hardness-ratio analysis}
\label{ssect:xmm_hrs}
\begin{figure}
\includegraphics[width=3.5in]{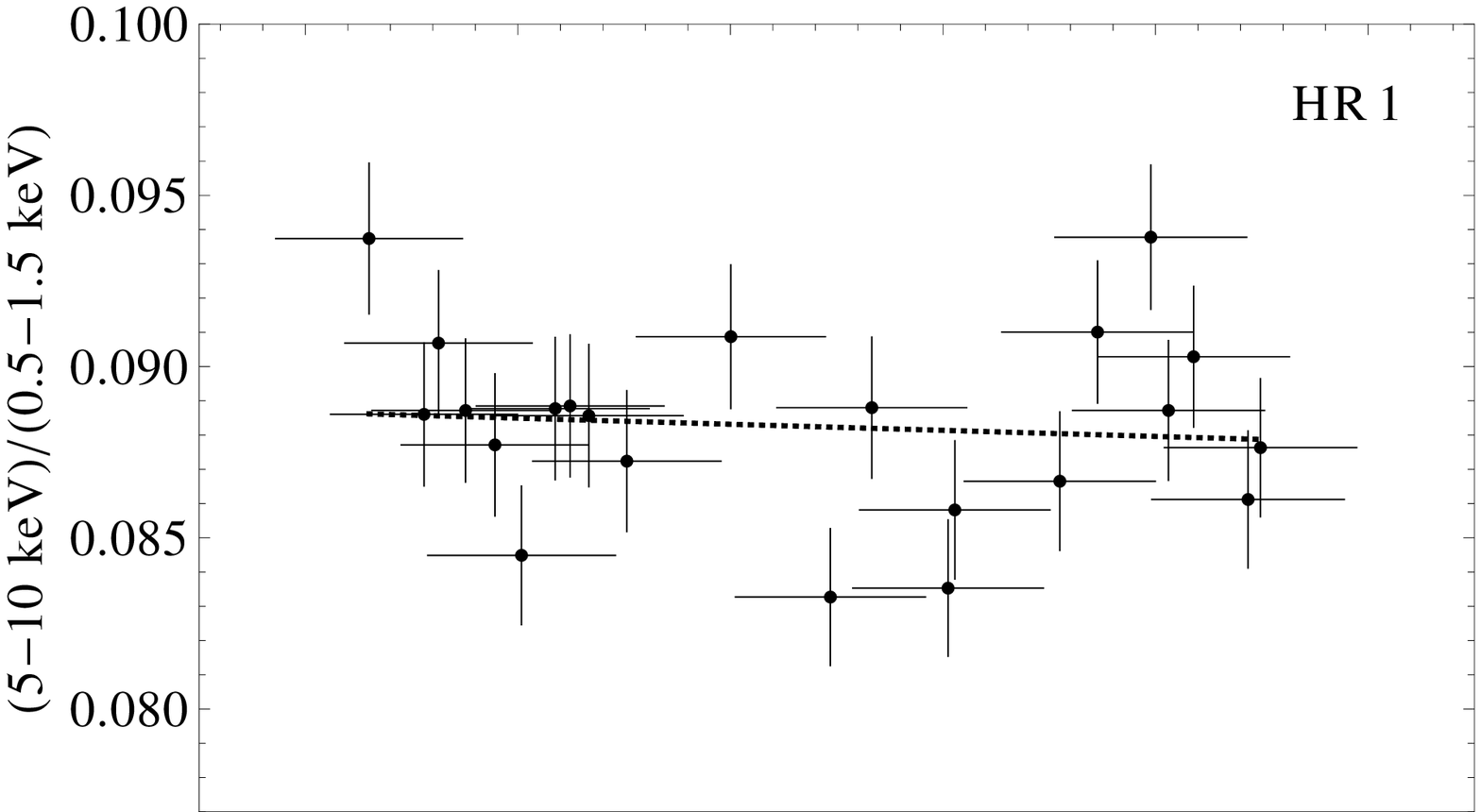}\\[-1.62em]
\hspace*{0.6em}\includegraphics[width=3.435in]{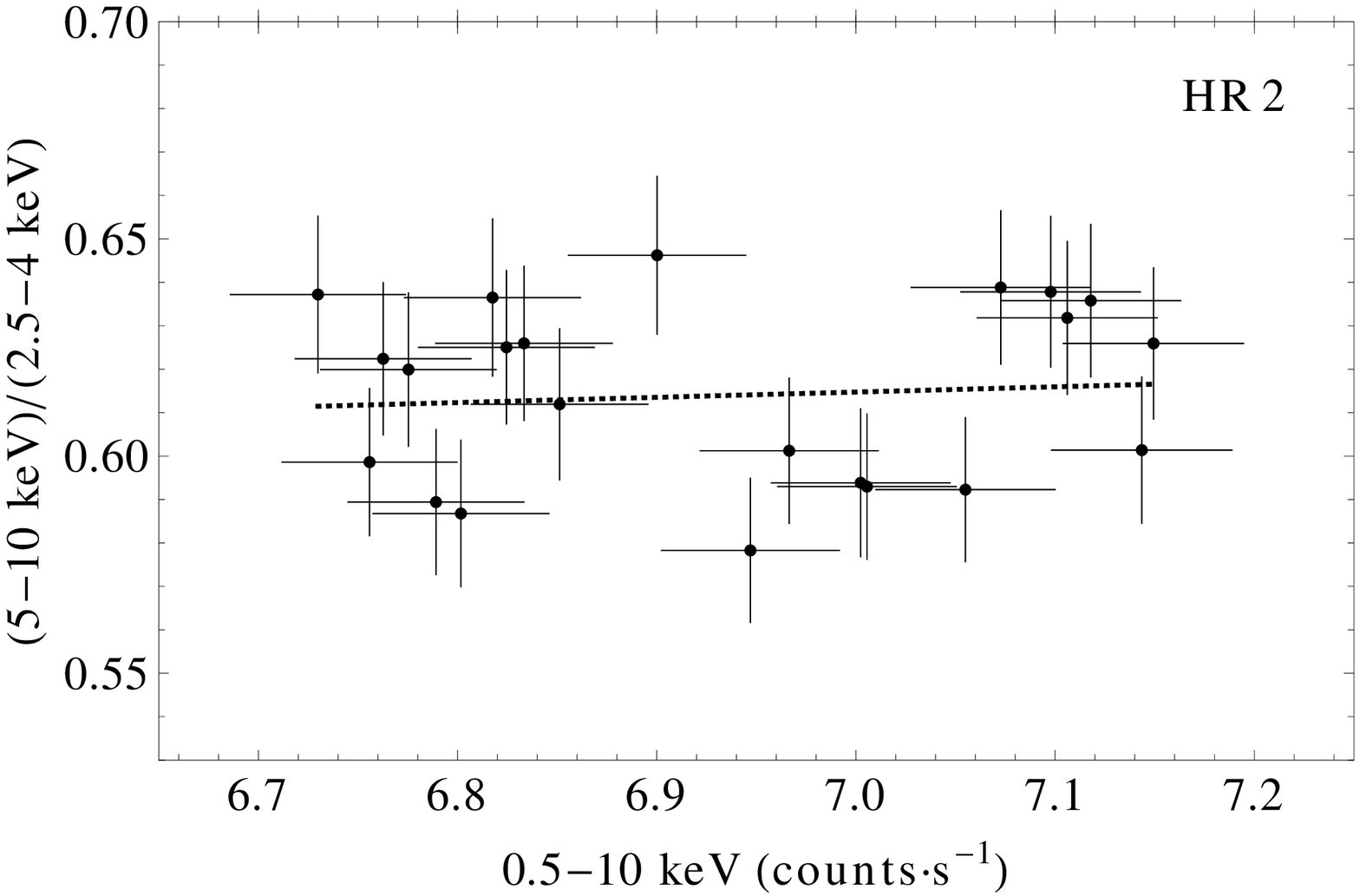}
\caption{Hardness-ratios in (5--10 keV)$/$(0.5--1.5 keV) and (5--10 keV)$/$(2.5--4 keV) versus the total count-rate in 0.5--10 keV, binned in 5 ks time-bins. The dotted-line represents the best-fitting linear model.}
\label{fig:xmm_hr_plot}
\end{figure}

After binning the light curves in bins of 5 ks, we estimate the hardness-ratios (5--10 keV)$/$(0.5--1.5 keV) (HR\,1) and (5--10 keV)$/$(2.5--4 keV) (HR\,2) versus the overall count-rate in 0.5--10 keV (Fig.~\ref{fig:xmm_hr_plot}). There is no direct evidence of an increasing or decreasing trend which would imply that the X-ray spectrum of the source becomes softer or harder when the source gets brighter. The error bars correspond to one standard deviation and they are derived following the standard error propagation procedure \citep[e.g.][]{bevington92}.\par
In order to check statistically HR\,1 for a potential trend, we fit to the 22 hardness-ratio points a linear model and we check how significantly the value of the slope differs from zero. For HR\,1 the best-fitting model gives a slope of $(-1.8\pm4.3)\times10^{-3}$ with a 95 per cent confidence interval (c.i.) of ($-1.1\times10^{-2}$,$7.2\times10^{-3}$)\footnote{The value of ${\itl t}$-statistic is ${\itl t}_{20,0.025}=2.09$.}. Since the value of ${\itl t}$-statistic that we get from the data-set is 3.37, the probability of getting such a value by chance alone is $3.05\times10^{-3}$. Similarly, for HR\,2 the best-fitting model gives a slope of $(1.2\pm3.2)\times10^{-2}$ having a 95 per cent c.i. of ($-5.5\times10^{-2},8.0\times10^{-2}$). The value of ${\itl t}$-statistic for this data-set is 2.36 yielding a chance coincidence probability of $2.8\times10^{-2}$.\par
These results imply that \n7 does not show significant X-ray spectral variations, as a function of time, and thus we can study the average overall X-ray spectra without being affected from temporal spectral variations.

\section{EPIC X-RAY SPECTRAL ANALYSIS}
\label{sect:epic_spec}
During the X-ray fitting procedures of the EPIC data we keep all the model-component parameters between the two MOS and the pn spectra tied together except from the normalisations (for the two MOS spectra they are still tied together), which are always left varying freely, unless otherwise stated. The model-flux is modified using the photoelectric absorption model {\tt wabs} having a fixed interstellar column-density equal to the measured hydrogen Galactic column of the neutral interstellar medium along this line of sight of $N_{\rm H}=1.1\times 10^{20}$ cm$^{-2}$ \citep[estimated using the {\sc ftool} \textit{nH}, after][]{kalberla05}.\par
Note, that the actual energies of the emission lines as well as the best-fit values for their centroid energies are given in the source's rest-frame. The horizontal axis in the spectral plots always refers to the energies in the observer's frame. For the X-ray spectral results, we use the $\chi^2$ fit statistic, with data variance, tracing the best-fit model parameters using the Levenberg-Marquardt algorithm \citep{bevington92}. The errors on the latter indicate their 90 per cent confidence range, corresponding to a $\Delta \chi^2$ of 2.706, unless otherwise stated. Finally, the best-fit parameters for the following X-ray spectral models, at 3--10 and 0.3--10 keV, are given in Table~\ref{tab:EPIC_spec_modelFits}.

\subsection{A first look at the broad-band X-ray spectrum}
\label{ssect:epic_broadBand}
\begin{figure}
\includegraphics[width=3.5in]{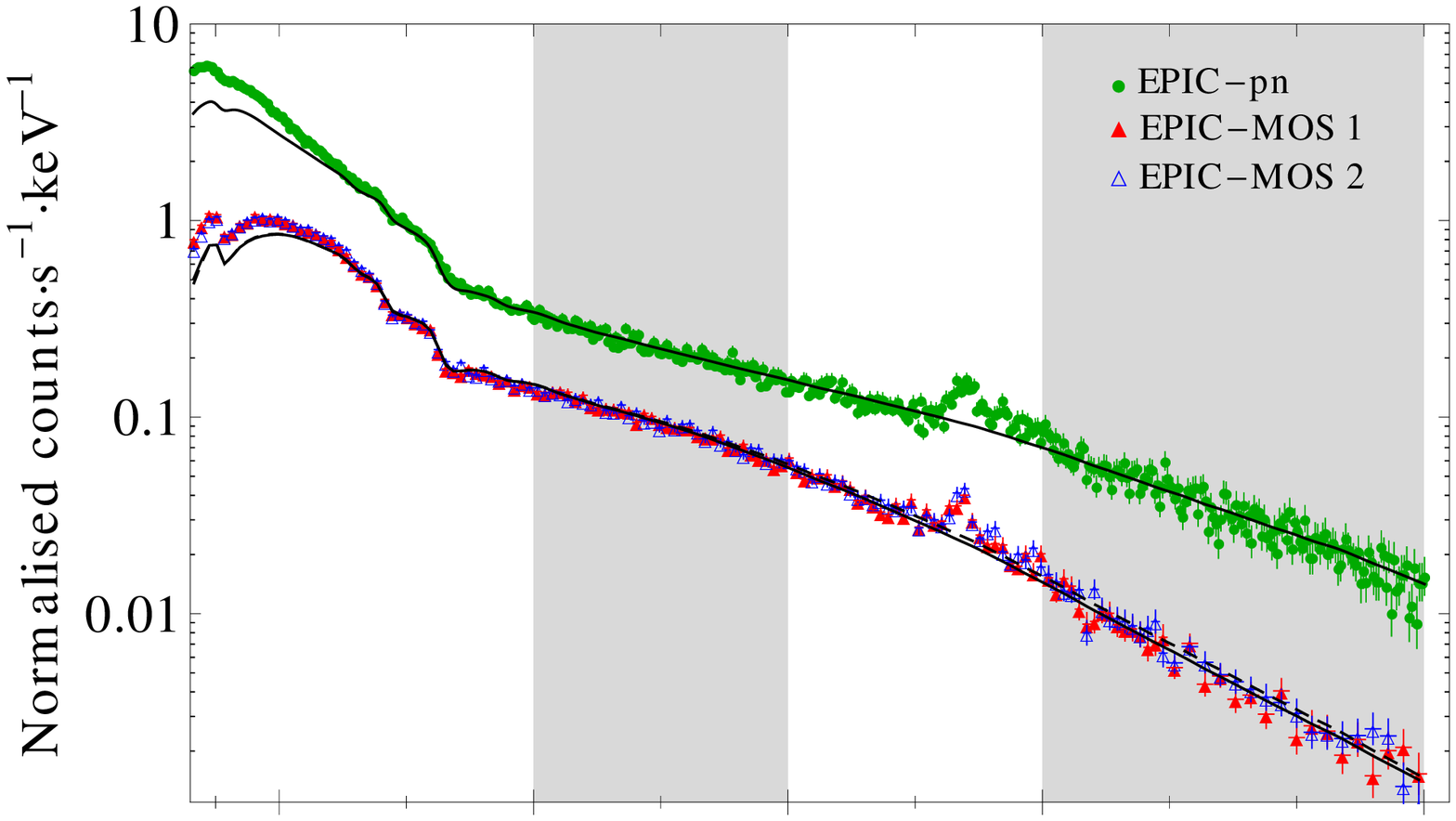}\\[-3.02em]
\hspace*{0.8em}\includegraphics[width=3.389in]{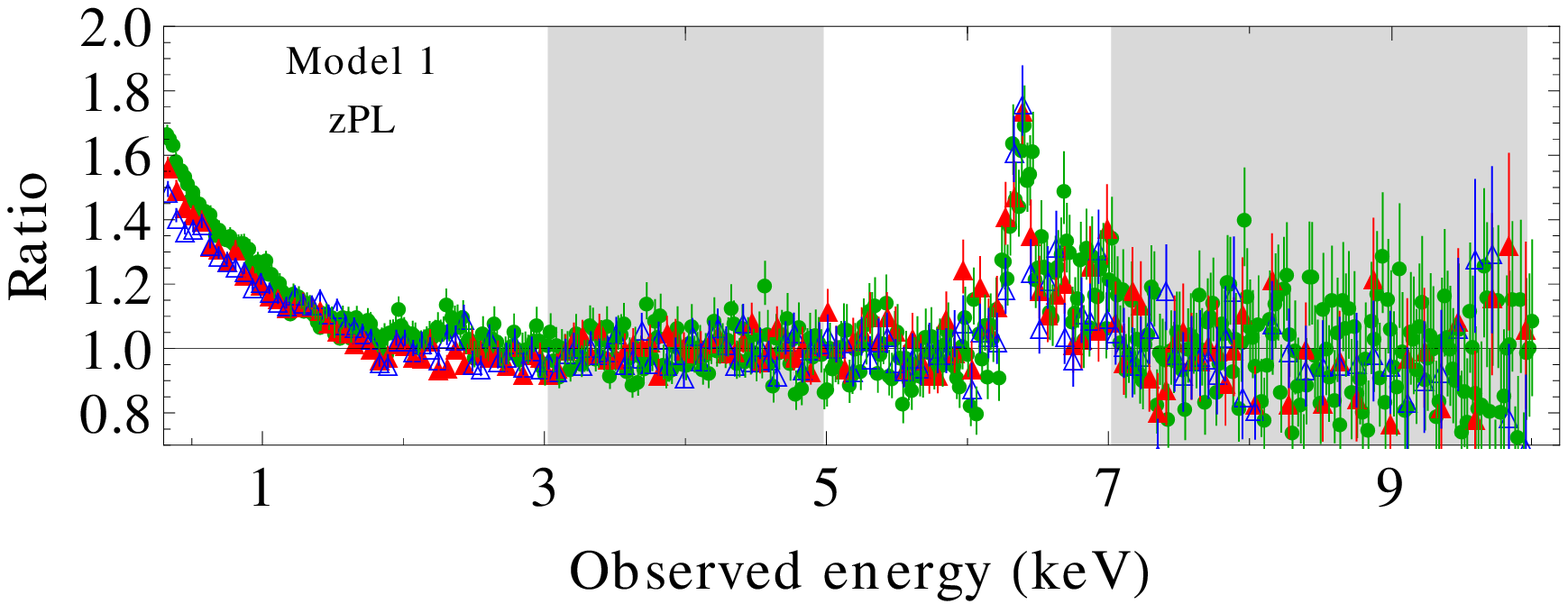}\\[-1.81em]
\caption{Model 1: Redshifted power-law fit in the observed 3--5 keV and 7--10 keV energy ranges, indicated by the grey-areas. The best-fit model is extrapolated over the energy range of 0.3--10 keV.}
\label{fig:spectral1}
\end{figure}

In order to get a phenomenological view of the overall EPIC X-ray spectra we initially fit a redshifted power-law (Model 1) to the data in the 3--5 keV and 7--10 keV energy-bands. The $\chi^2$ value of the best-fit model is equal to 404.57 for 387 d.o.f. ($0.259$) null-hypothesis probability (NHP)) yielding a best-fit photon spectral index of $\Gamma=1.66\pm0.02$. Figure~\ref{fig:spectral1} shows this best-fit model extrapolated over the energy range of 0.3--10 keV together with the respective ratio plot.

From the ratio plot of Fig.~\ref{fig:spectral1} we can readily distinguish three very distinct spectral features. Firstly, below 2 keV, we notice a flux excess on top of the best-fitting power-law extrapolation. Then, we can discern a strong emission line appearing around 6.36 keV and finally several excess features between 6.4 and 7.2 keV.\par
\begin{figure}
\includegraphics[width=3.397in]{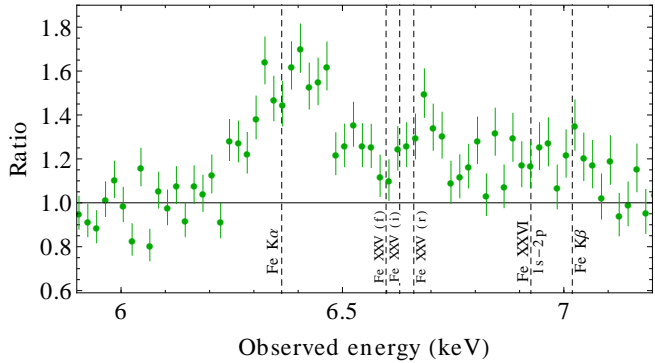}
\caption{The EPIC-pn spectral data between 5.9 and 7.2 keV together with the expected positions of some Fe species indicated by the vertical dashed lines.}
\label{fig:spectral1_FeRegion}
\end{figure}

For clarity reasons, in Fig.~\ref{fig:spectral1_FeRegion} we show a blow-up of the same ratio plot, for the region 5.9--7.2 keV. We plot only the ratio points of the pn data, together with the position of the following iron emission lines that are frequently observed/assumed in the AGN X-ray spectra: a neutral \fa (6.4 keV), He-like Fe {\sc xxv}-[forbidden (f), intercombination (i) and resonance (r)] (6.637, 6.668, and 6.700 keV), H-like Fe {\sc xxvi} [1s--2p] (6.966 keV) and a neutral \fb (7.06 keV).

\subsection{Spectral fitting in the 3--10 keV energy range}
\label{ssect:epic_3_10}
Since Model 1 (Fig.~\ref{fig:spectral1}) gives an adequate description of the data in the 3--5 keV and 7--10 keV energy ranges, we keep it as a baseline model, for the following spectral analysis over the 3--10 keV energy range. As we discuss below, the excess emission between 6.4 and 7.2 keV is modelled as an ensemble of individual narrow Gaussian function having fixed the standard deviation at 22 eV. This value corresponds to a full width half maximum (FWHM) of 2400 km s$^{-1}$ \citep{bianchi08,shu10} and it is well below the actual energy resolution of the EPIC instruments (i.e.\ around 150 eV FWHM at 6.4 keV for both MOS and pn).

\subsubsection{Iron line species}
\label{ssect:epic_iron_line_comples}
Initially, in order to account for the observed emission line at 6.37 keV, which appears to correspond to the \fa line from neutral iron (having a centroid energy of 6.4 keV), we consider a redshifted narrow Gaussian function. For consistency reasons with the theoretical predictions \citep{kaastra93}, the presence of a neutral \fa requires the existence of an associated \fb line. Therefore, we add another redshifted narrow Gaussian function to account for the \fb line, fixing its centroid energy at 7.06 keV and its normalisation at 14 per cent of that of the \fa's normalisation value \citep{molendi03}. Using \textit{Chandra}-HETG, \citet{bianchi09} found no evidence of a \fb line yielding a 90 per cent upper confidence limit for the for its normalisation of 18 per cent of the \fa's normalization, being consistent with the theoretically expected value. However, the presence of a narrow K$\beta$ line is not statistically required from our spectral data (i.e.\ its addition to the models described below does not significantly improve the quality of the model fits). However, for physical reasons, given the presence of the strong neutral iron K$\alpha$ emission line in the spectrum of the source, we decide to keep this emission line in our X-ray spectral fits (note that we verified that our model fit results are not significantly altered if we exclude this component).\par
\begin{figure}
\hspace*{0.01em}\includegraphics[width=3.503in]{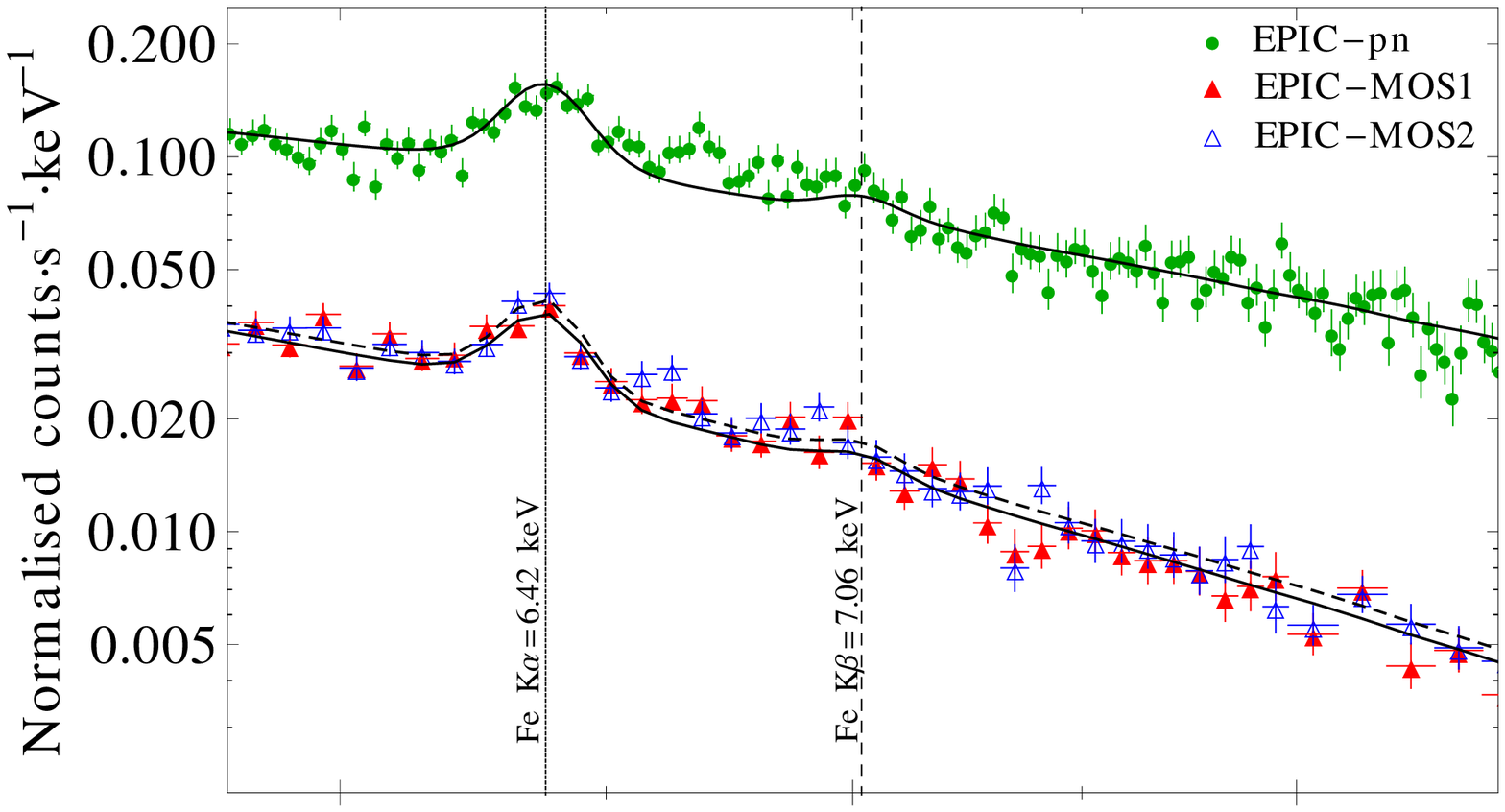}\\[-2.19em]
\hspace*{1.46em}\includegraphics[width=3.315in]{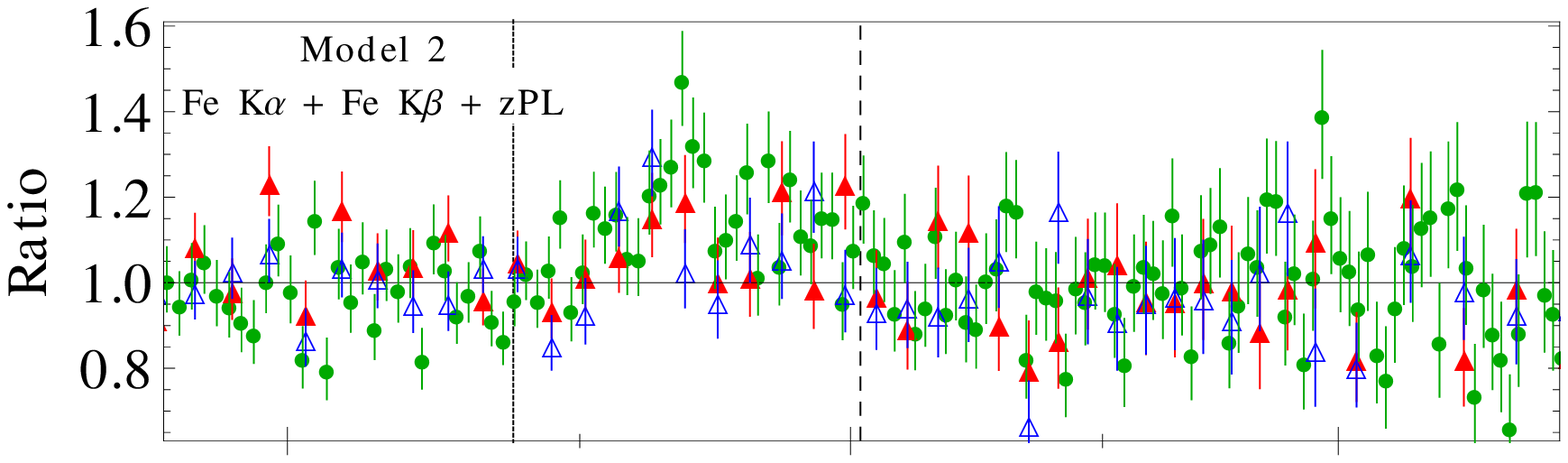}\\[-2.91em]
\hspace*{0.01em}\includegraphics[width=3.504in]{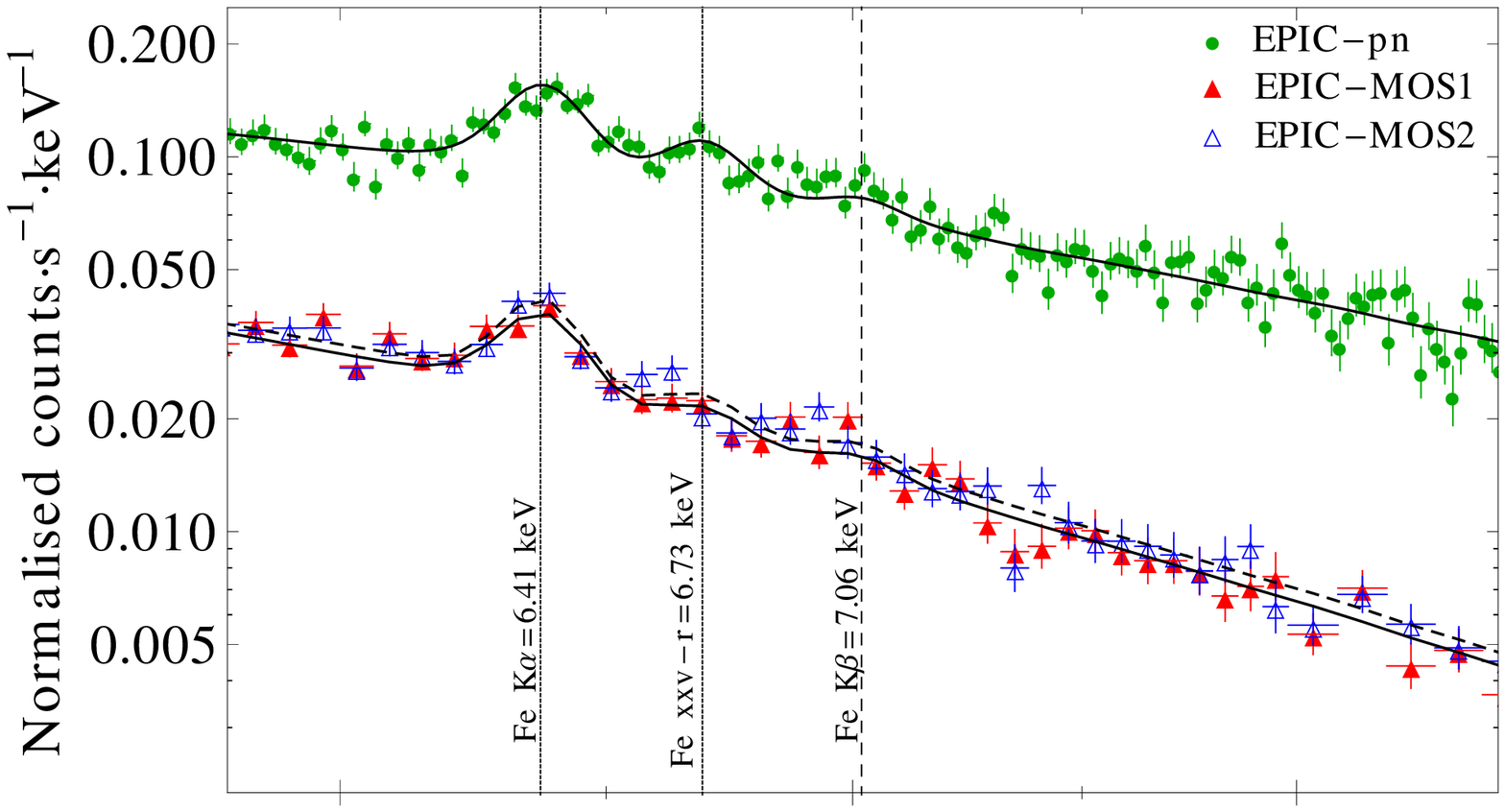}\\[-2.17em]
\hspace*{1.46em}\includegraphics[width=3.314in]{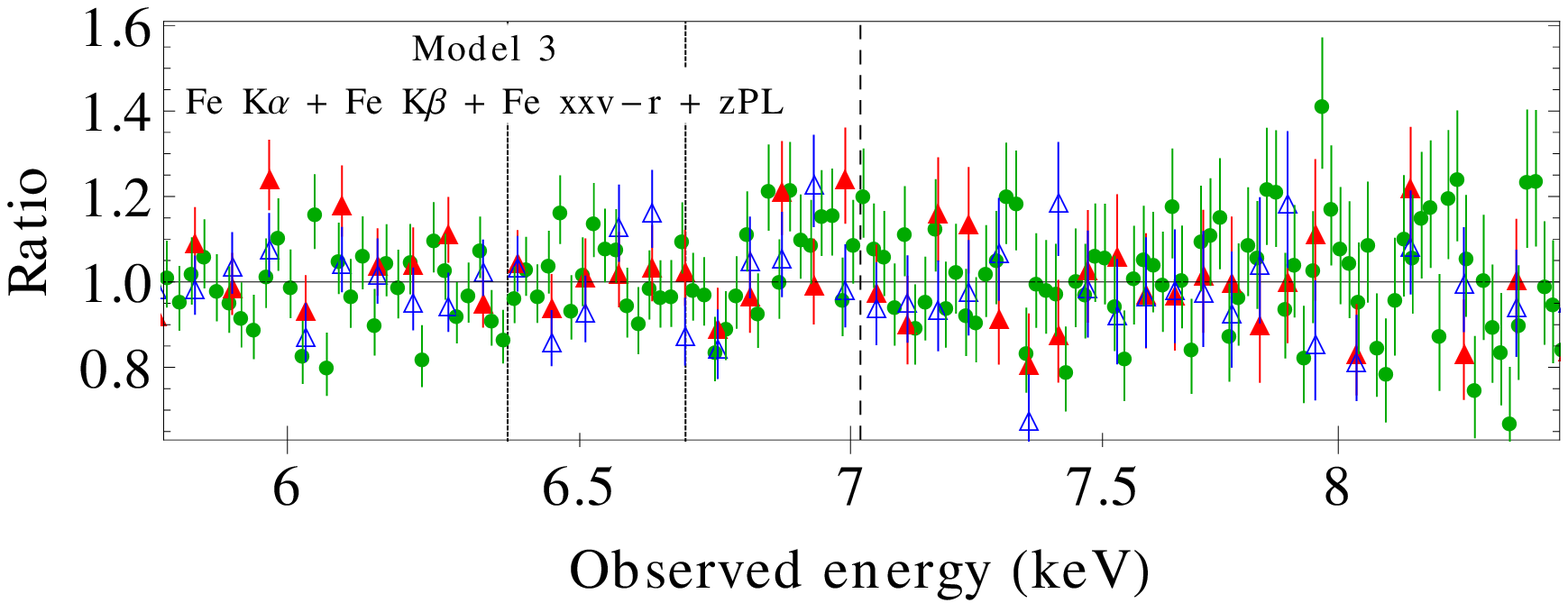}
\caption{Line spectral fits in the observed 3--10 keV energy range. The vertical labels indicate the energy in the source's rest-frame. (Top panel) Model 2: Two Fe lines [\fa (free), \fb (fixed)];a redshifted power-law. (Bottom panel) Model 3: Three Fe lines [\fa (free), Fe {\sc xxv}-r (free), \fb (fixed)];a redshifted power-law.}
\label{fig:spectral2_3}
\end{figure}

The best-fitting model (Model 2, Fig.~\ref{fig:spectral2_3}, top panel) has a $\chi^2$ value of 696.13 for 547 d.o.f. ($1.5\times10^{-5}$ NHP) and its best-fitting values for the Fe K$\alpha$'s centroid energy and the power-law's photon index are $6.42\pm0.01$ keV and $1.63\pm0.02$, respectively. From the corresponding ratio plot (Fig.~\ref{fig:spectral2_3}, top panel) we can readily see that the ratios around 6.37 keV and 7.02 keV are consistent with unity but there is still a great deal of spectral excess between the observed energy range of 6.5--7 keV.\par
In order to model this excess we consider additionally a third redshifted narrow Gaussian function (Model 3). The best-fitting Model 3 gives a considerably better $\chi^2$ value of 617.05 for 544 d.o.f. ($1.6\times10^{-2}$ NHP) yielding for the new component a best-fitting centroid energy of $6.73^{+0.01}_{-0.05}$ keV. The centroid energy of the \fa line as well as the photon index of the power-law are still consistent with the previously derived values (i.e.\ Model 2). The bottom panel of Fig.~\ref{fig:spectral2_3} shows the best-fitting model together with the corresponding ratio plot, the latter, showing an excess around 6.8 keV.\par
\begin{figure}
\hspace*{-0.06em}\includegraphics[width=3.510in]{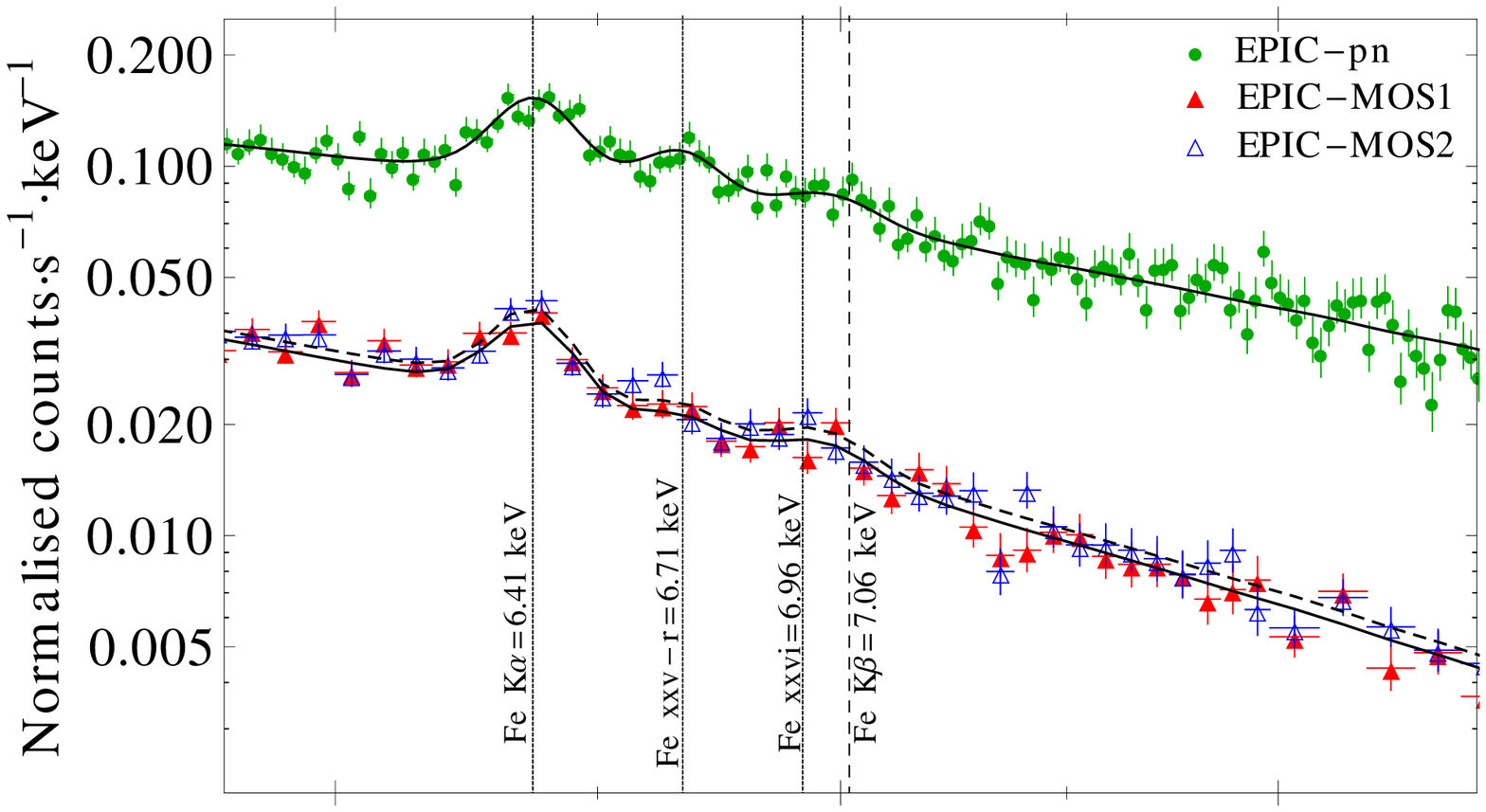}\\[-2.18em]
\hspace*{1.39em}\includegraphics[width=3.320in]{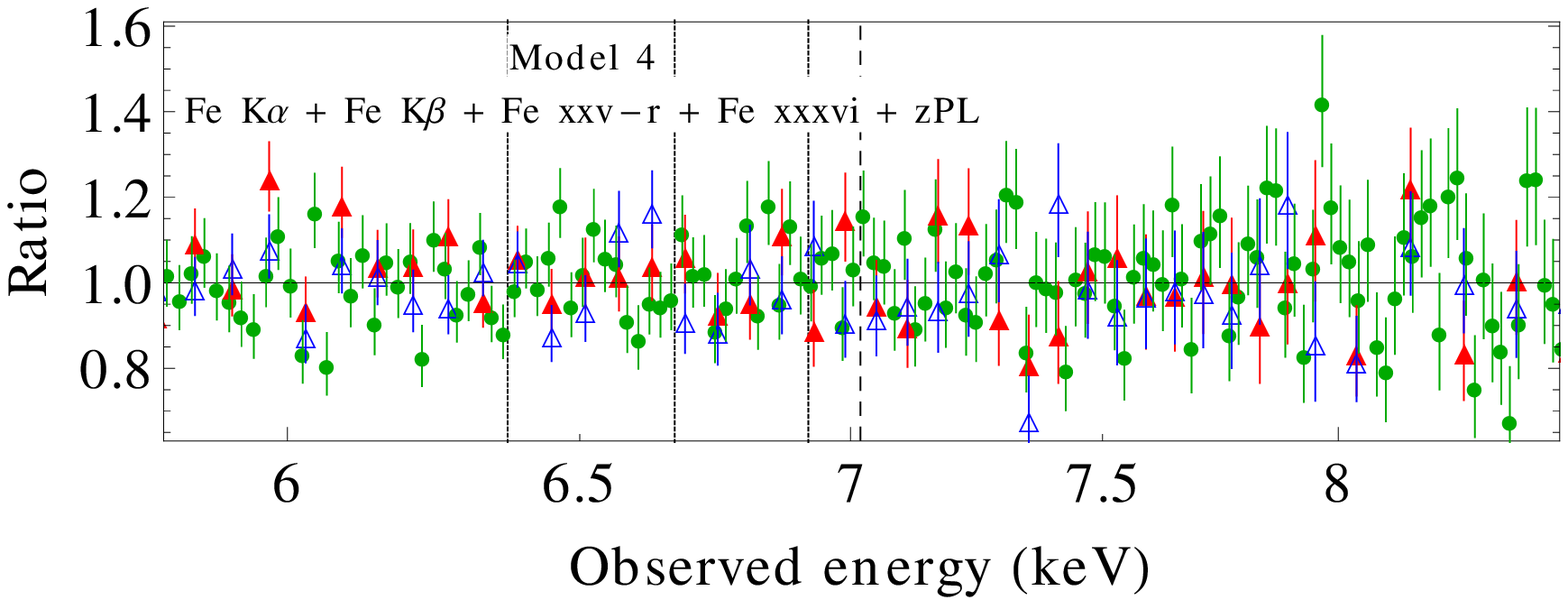}
\caption{Line spectral fits in the observed 3--10 keV energy range for Model 4. The vertical labels indicate the energy in the source's rest-frame. Four Fe lines [\fa (free), Fe {\sc xxv}-r (free), Fe {\sc xxvi} (free), \fb (fixed)];a redshifted power-law.}
\label{fig:spectral4}
\end{figure}

In order to model this excess we add a fourth redshifted narrow Gaussian function (Model 4, Fig.~\ref{fig:spectral4}). The addition of this narrow line component ameliorates significantly the $\chi^2$ value of the best-fitting Model 4, 593.41 for 541 d.o.f. (0.06 NHP), yielding a best-fitting centroid energy of $6.96^{+0.03}_{-0.05}$ keV. The best-fitting centroid energy of the third narrow emission line (i.e.\ the one introduced in Model 3) is now $6.71\pm0.02$ keV and the best-fitting values for the Fe K$\alpha$'s centroid energy and power-law photon index remain exactly the same with those ones derived previously from the best-fitting Model 3.\par
\begin{figure}
\includegraphics[width=3.397in]{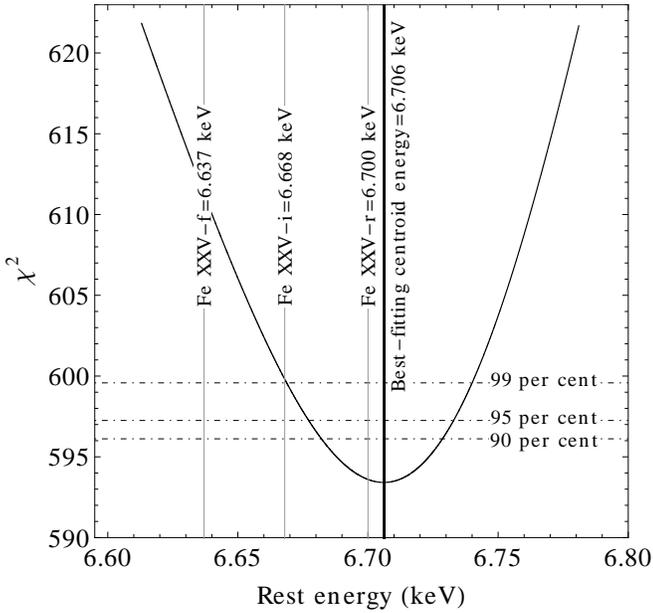}
\caption{The $\chi^2$ versus the centroid rest energy of the 6.71 keV emission line of Model 4. The thin vertical grey lines indicate the rest energies of the Fe {\sc xxv} triplet components. The horizontal dot-dashed lines indicate the 90, 95 and 99 per cent confidence ranges.}
\label{fig:FeXXVtriplet_chiSquare}
\end{figure}

The narrow emission line, having a best-fitting centroid energy of 6.96 keV, should correspond to the 1s--2p doublet from H-like Fe {\sc xxvi}. The other emission line, having a best-fitting centroid energy of 6.71 keV, is consistent with the He-like Fe {\sc xxv}-r since its 99 per cent c.i. is (6.667--6.741) keV (Fig.~\ref{fig:FeXXVtriplet_chiSquare}). 
Using the best-fitting results of Model 4, we can now estimate the equivalent widths (EWs) for the various iron line species, having as an underlying continuum the power-law with photon index $1.66$. For the \fa and the \fb we get $126^{+32}_{-27}$ and $23^{+6}_{-5}$ eV respectively, and for the Fe {\sc xxvi} and Fe {\sc xxv}-r we get $23^{+29}_{-23}$ and $58^{+31}_{-20}$ eV respectively. 

\subsubsection{Origin of \fa line}
\label{ssect:epic_originFaLine}
Next we consider whether the existence of a distant Compton-thick reflector can further improve the fit providing at the same time a natural explanation for the \fa line. To test that we fix all the iron line components to their best-fitting values from Model 4, apart from that of \fa line, and we include to that the additive {\sc xspec} model {\tt pexrav} \citep{magdziarz95}. 
During the fit, we tie the photon index of the {\tt pexrav}-model to that of the power-law continuum, we fix the abundance of elements heavier than helium to that of the solar value and we fix the cut-off energy to 1000 keV. Then, we consider an ensemble of inclinations $30\degr$, $50\degr$ and $80\degr$ as well as iron abundances 0.6, 0.8 and 1 and we estimate the best-fitting reflection scaling factor $R$ for each case. For all the cases, the best-fit {\tt pexrav} normalisation is equal to zero having a $\chi^2$ value of 589.65 for 541 d.o.f. (0.07 NHP). The reduced $\chi^2$s between this model and Model 4 (1.09 and 1.10 respectively) imply that such a Compton-thick distant reflection component is not required from our spectral data.\par 
Note that by excluding all the iron line species (i.e.\ Gaussian components) and fit the spectral data, with the above-mentioned set of {\tt pexrav} models, we get as a best fit model the one having a frozen inclination of $50\degr$ and an iron abundance of 0.8. This model yields a $\chi^2$ value of 1009.11 for 548 d.o.f. (0 NHP) being worse than that of Model 4 and thus making the presence of the iron line components statistically more necessary than that of a Compton-thick distant reflection component. Therefore, in the following broad-band spectral fits we exclude the {\tt pexrav}-model from our analysis.

\subsection{The 0.3--10 keV EPIC-pn spectrum: Thermal plasma components}
\label{ssect:epic_broad_band_Spec}
As we have seen from the ratio plot of Fig.~\ref{fig:spectral1}, \n7 exhibits below 2 keV an excess flux (i.e.\ additional to the power-law component) known as `soft excess' emission.\par
\begin{figure}
\hspace*{-0.1em}\includegraphics[width=3.515in]{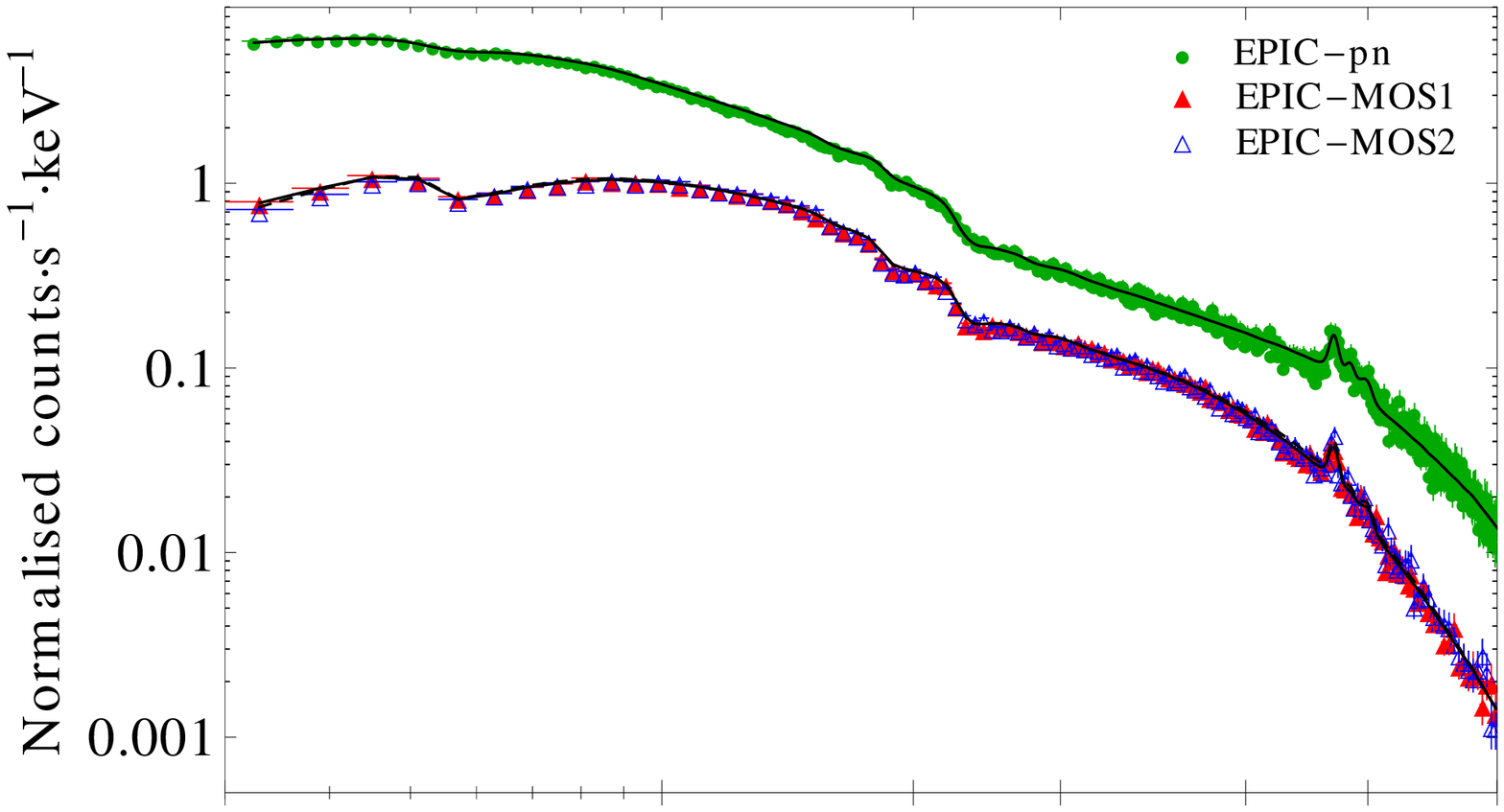}\\[-2.26em]
\hspace*{1.31em}\includegraphics[width=3.330in]{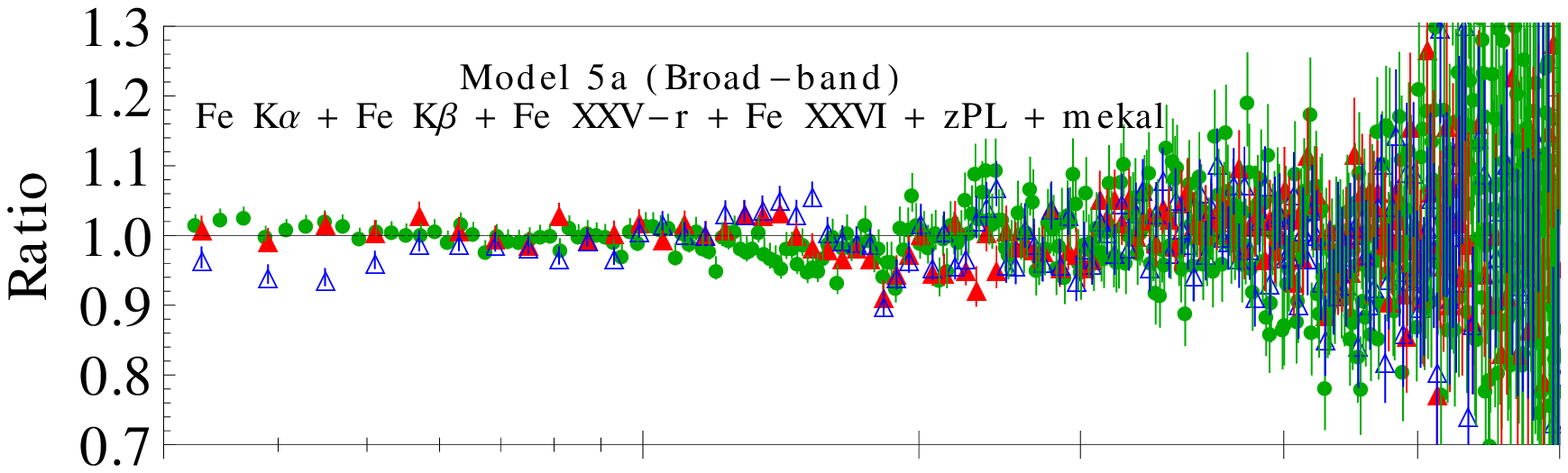}\\[-2.94em]
\hspace*{-0.1em}\includegraphics[width=3.512in]{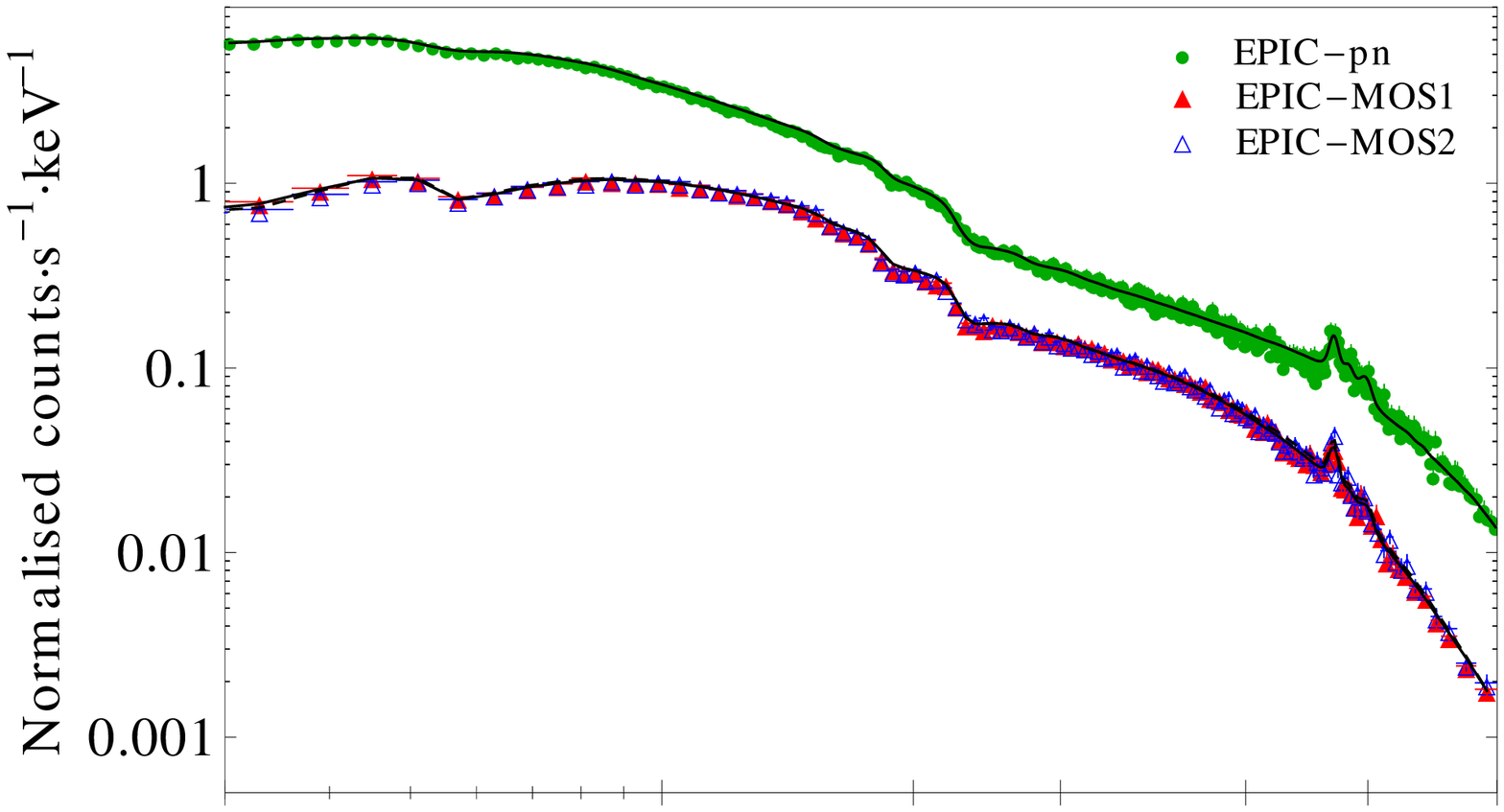}\\[-2.27em]
\hspace*{1.25em}\includegraphics[width=3.401in]{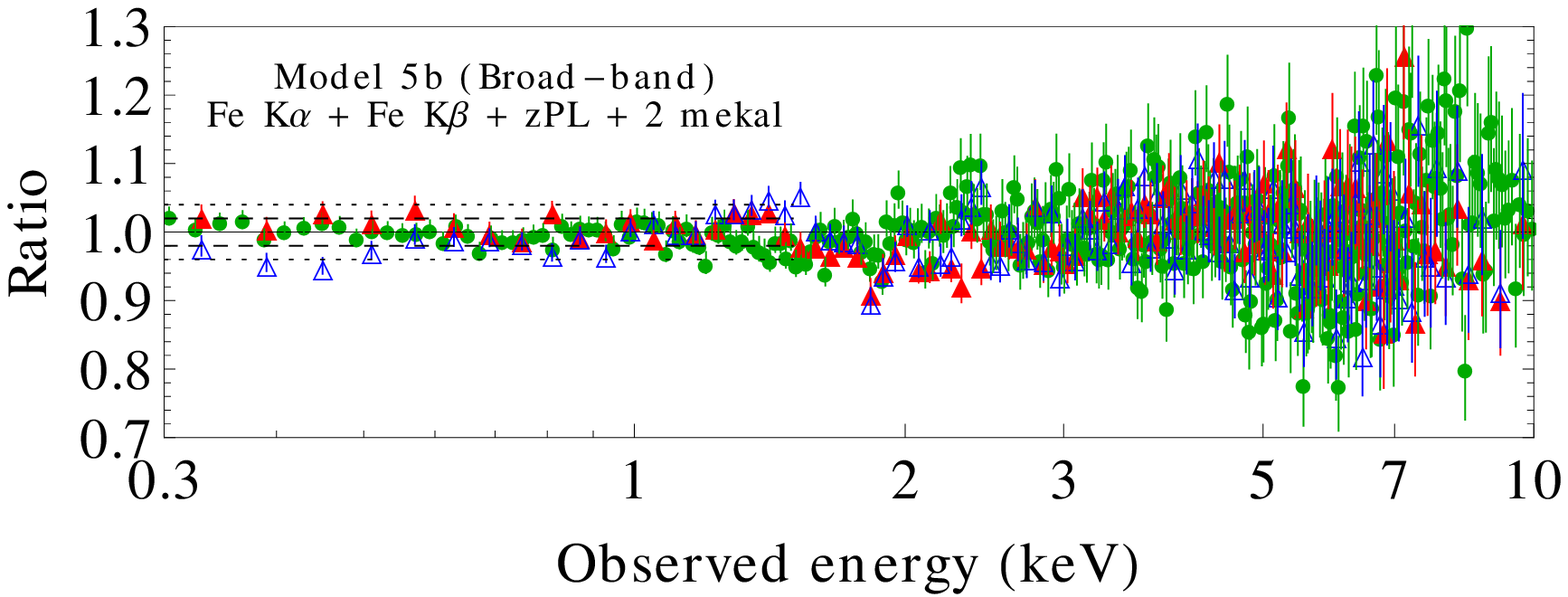}
\caption{EPIC spectral fits in the observed 0.3--10 keV energy range. (Top panel) Model 5a: Four Fe lines [\fa (fixed), Fe {\sc xxv}-r (fixed), Fe {\sc xxvi} (fixed), \fb (fixed)]; a redshifted power-law; a thermal component. (Bottom panel) Model 5b: Four Fe lines [\fa (fixed), Fe {\sc xxv}-r (fixed), Fe {\sc xxvi} (fixed), \fb (fixed)]; a redshifted power-law; two thermal components. The dotted and dashed lines correspond to the 2 and 4 per cent cross calibration uncertainties.}
\label{fig:spectral5a_b}
\end{figure}

In order to fit the `soft excess' we extend the best-fit Model 5 down to 0.3 keV and we include the additive \textit{XSPEC} thermal model {\tt mekal} \citep{mewe85,raymond09} incorporating the emission spectra from hot diffuse gas (thermal plasma). During the fit we fix the energies of the iron line species to their best fitting values as derived from Model 4 and for the {\tt mekal}-model we freeze both the Hydrogen volume density and the chemical abundance of metals to unity and to that of Solar abundance, respectively. The best-fitting model (Model 5a, Fig.~\ref{fig:spectral5a_b}, top panel) has a $\chi^2$ of 896.44 for 764 d.o.f. ($6.26\times10^{-4}$ NHP) yielding a best-fitting plasma temperature of $k_{\rm B}T=0.34\pm0.04$ keV and a best-fitting power-law photon index of $1.86^{+0.02}_{-0.03}$.\par

\begin{table*}
\begin{minipage}{190mm}
\caption{Best-fit spectral results for the EPIC science products.}
\label{tab:EPIC_spec_modelFits}
\begin{tabular}{@{}lccccccc}
\hline
\multirow{2}{*}{Model-parameter}  & \multirow{2}{*}{Model 1} & \multirow{2}{*}{Model 2} & \multirow{2}{*}{Model 3} & \multirow{2}{*}{Model 4} & \multirow{2}{*}{Model 5a} & \multirow{2}{*}{Model 5b} \\[0.4em]
 & [(3--5) $\cup$ (7--10)] keV &  3--10 keV & 3--10 keV & 3--10 keV & 0.3--10 keV  & 0.3--10 keV \\
\hline
1a) {\tt zpowerlw}: Photon index, $\Gamma$ & $1.66\pm0.02$ & $1.63\pm0.02$ & $1.66^{+0.01}_{-0.02}$ & $1.66^{+0.01}_{-0.02}$ &   $1.86^{+0.02}_{-0.03}$ &  $1.86\pm0.02$ \\
1b) {\tt zpowerlw}: Normalisation\textsuperscript{1} & $2.84\pm0.01$ & $2.78\pm0.02$ & $2.87\pm0.02$ & $2.88\pm0.02$ & $3.27\pm0.02$ & $3.15\pm0.02$\\
2a) {\tt zGauss}: \fa line-energy (keV) &  & $6.42\pm0.01$ & $6.41\pm0.01$ &  $6.41\pm0.01$ & 6.41 & 6.41 \\
2b) {\tt zGauss}: Normalisation\textsuperscript{2} & & $1.82^{+0.16}_{-0.20}$ & $1.85^{+0.16}_{-0.19}$ & $1.78^{+0.15}_{-0.16}$ & $1.55^{+0.13}_{-0.16}$ & $1.48\pm0.14$\\
3a) {\tt zGauss}: \fb line-energy (keV) &  & 7.06  & 7.06 & 7.06 & 7.06 & 7.06\\
3b) {\tt zGauss}: Normalisation\textsuperscript{2} &  & 0.14$\times$2b\textsuperscript{\textasteriskcentered}  & 0.14$\times$2b\textsuperscript{\textasteriskcentered} & 0.14$\times$2b\textsuperscript{\textasteriskcentered}  &  0.14$\times$2b\textsuperscript{\textasteriskcentered} & 0.14$\times$2b\textsuperscript{\textasteriskcentered} \\
4a) {\tt zGauss}: Fe {\sc xxv}-r line-energy (keV) &  &  & $6.73^{+0.01}_{-0.05}$ & $6.71\pm0.02$ & 6.71 & ---\\
4b) {\tt zGauss}: Normalisation\textsuperscript{2} &  & & 7.3$^{+1.2}_{-2.6}$ & 5.6$^{+2.1}_{-1.5}$ & 4.67$^{+1.1}_{-2.7}$ & ---\\
5a) {\tt zGauss}: Fe {\sc xxvi} line-energy (keV)  &  & & & $6.96^{+0.03}_{-0.05}$ & 6.96 & ---\\
5b) {\tt zGauss}: Normalisation\textsuperscript{2} &  & &  & $3.0^{+2.1}_{-1.9}$ & $1.3^{+3.2}_{-0.8}$ & ---\\
6a) {\tt mekal}: Temperature, $k_{\rm B}T$ (keV) & & & & & $0.34\pm0.04$ & $0.36\pm0.04$\\
6b) {\tt mekal}: H density, $n_{\rm H}$ (cm$^{-3}$) & & & & & 1 & 1\\
6c) {\tt mekal}: Metal abundance & & & & & 1 & 1\\
6d) {\tt mekal}: Normalisation\textsuperscript{3}   & & & & & $4.72^{+1.68}_{-1.30}$ & $4.58^{+1.28}_{-1.42}$\\
7a) {\tt mekal}: Temperature, $k_{\rm B}T$ (keV) & & & & & & $8.84^{+1.78}_{-1.11}$ \\
7b) {\tt mekal}: H density, $n_{\rm H}$ (cm$^{-3}$) & & & & & & 1\\
7c) {\tt mekal}: Metal abundance & & & & & & 1 \\
7d) {\tt mekal}: Normalisation\textsuperscript{3}  & & & & & & $50.2^{+24.1}_{-14.2}$\\
\hline
$\chi^2/$d.o.f. & $404.57/387$ & $696.13/547$ & $617.05/544$ & $593.41/541$  & $896.44/764$ & $900.70/767$ \\
\hline
\end{tabular}
\medskip
\\
Values without uncertainties correspond to fixed fitting parameters.\\ The best-fitting normalisations correspond to the average of the two MOS (tied together) and pn.\\
\textsuperscript{\textasteriskcentered} 14 per cent of the \fa line's normalisation value (parameter 2b).\\
\textsuperscript{1} In units of 10$^{-3}$ phot. keV$^{-1}$cm$^{-2}$ s$^{-1}$ at 1 keV.\\
\textsuperscript{2} In units of 10$^{-5}$ phot. cm$^{-2}$ s$^{-1}$ at the line energy.\\
\textsuperscript{3} In units of $\int n_{\rm e}\,n_{\rm H} dV$, ($10^{62}$ cm$^{-3}$).\\
\end{minipage}
\end{table*}

Finally, we try to model the 0.3--10 keV X-ray spectrum in an entirely self consistent way. Therefore, we remove the two iron line species at 6.71 and 6.96 keV, represented by the two Gaussian components, and we add a second thermal component (Model 5b, Fig.~\ref{fig:spectral5a_b}, bottom panel), assuming in this way that these iron species originate from the emission spectrum of a hot diffuse gas. During the fit the normalisations of the various model components are left to vary freely. The $\chi^2$ of the fit is similar to that of Model 5a, being 900.70 for 767 d.o.f. ($5.77\times10^{-4}$ NHP) yielding a low plasma temperature of $k_{\rm B}T_1=0.36\pm0.04$, a high plasma temperature of $k_{\rm B}T_{2}= 8.84^{+1.78}_{-1.11}$ keV and a power-law index of $1.86\pm0.02$. Note that this model is more reliable not only because it yields a marginally better NHP than Model 5a, but it also provide us a physical explanation about the origin of the ionised iron line species.\par
The best-fit results are plotted in the bottom panel of Fig.~\ref{fig:spectral5a_b} having no significant residuals remaining in the 0.3--10 keV energy band. The relatively poor value of $\chi^2$ is mainly due to the residuals, below 1.5 keV. The ratio plot, indicates that the pn, MOS\,1 and MOS\,2 residuals are consistent with each other at energies larger than 0.5 keV, and their amplitude is mostly smaller than 2 per cent in the whole energy range between 0.3 and 1.5 keV (and up to 4 per cent in a few cases; the dashed an dotted lines in this plot indicate the 2 and 4 per cent amplitudes, respectively, in the 0.3-1.5 keV band). On the other hand, the MOS\,2 residuals do show discrepancies with the best-fit model ({\it and} the pn and MOS1) of the order of 6 per cent at energies below 0.5 keV, and in fact they contribute with a significant part to the resulting $\chi^2$. Given the fact that the best model-fit residuals do not exhibit well structured features (indicative of the presence of an extra spectral component that we have not taken into account), we conclude that the low amplitude soft energy band residuals may be due to remaining calibration uncertainties of each detector, and also un-modelled cross-calibration uncertainties between the individual detectors as well\footnote{For calibration related issued read the presentation of M.Guainazzi at \url{http://xmm.esac.esa.int/external/xmm_user_support/usersgroup/20120419/index.shtml}.}.\par
Finally, we also investigate whether the addition of a second, low energy thermal component could improve the quality of the model fit (previous X-ray spectra of the source have been modelled by two thermal components in the soft energy band; see the Discussion section). However, the addition of this spectral component does not improve the quality of the fit, since it yields a $\chi^2$ of 899.12 for 764 dof (5.04$\times 10^{-4}$ NHP). In fact, the best-fit temperature of the second component is consistent with zero (with a 90 per cent upper confidence limit of 9.7$\times 10^{-2}$ keV). This was the case both when we leave the temperature of the first thermal component and the power-law spectral slope to vary freely during the model fit, and when we keep them frozen to their best-fit values of the previous model fit.\par
After estimating the best-fitting model parameters of Model 5b for the overall spectrum, we can compute the X-ray flux and X-ray luminosity of \n7. In the 0.5--10 keV and 2--10 keV energy ranges the absorbed[unabsorbed] flux is $(1.92[1.99]\pm0.01)\times10^{-11}$ erg cm$^{-2}$ s$^{-1}$ and $(1.22[1.23]\pm0.01)\times10^{-11}$ erg cm$^{-2}$ s$^{-1}$, respectively. The corresponding X-ray luminosities for the same energy bands are $(1.12[1.16]\pm0.02)\times10^{42}$erg s$^{-1}$ and $(7.14[7.20]\pm0.03)\times10^{41}$erg s$^{-1}$, respectively (the errors in the fluxes and the luminosities indicate the 68 per cent c.i. in order to be comparable with previous observations). Using the 2--10 keV luminosity, we can estimate the expected EW of \fa using eq.\ 2 in \citet{bianchi09}. This yields a value of around 124 eV being very consistent with our estimate, being around 126 eV (Section \ref{ssect:epic_iron_line_comples}).

\section{RGS X-RAY SPECTRAL ANALYSIS}
\label{sect:rgs_spec}
We first model the 0.35--1.8 keV RGS spectra of \n7 independently on the results of the EPIC spectral fitting, in the overlapping energy ranges (i.e.\ above 0.4  keV). The best-fit parameters for the following X-ray spectral models are given in Table~\ref{tab:RGS_spec_modelFits}. \par
\begin{figure}
\includegraphics[width=3.503in]{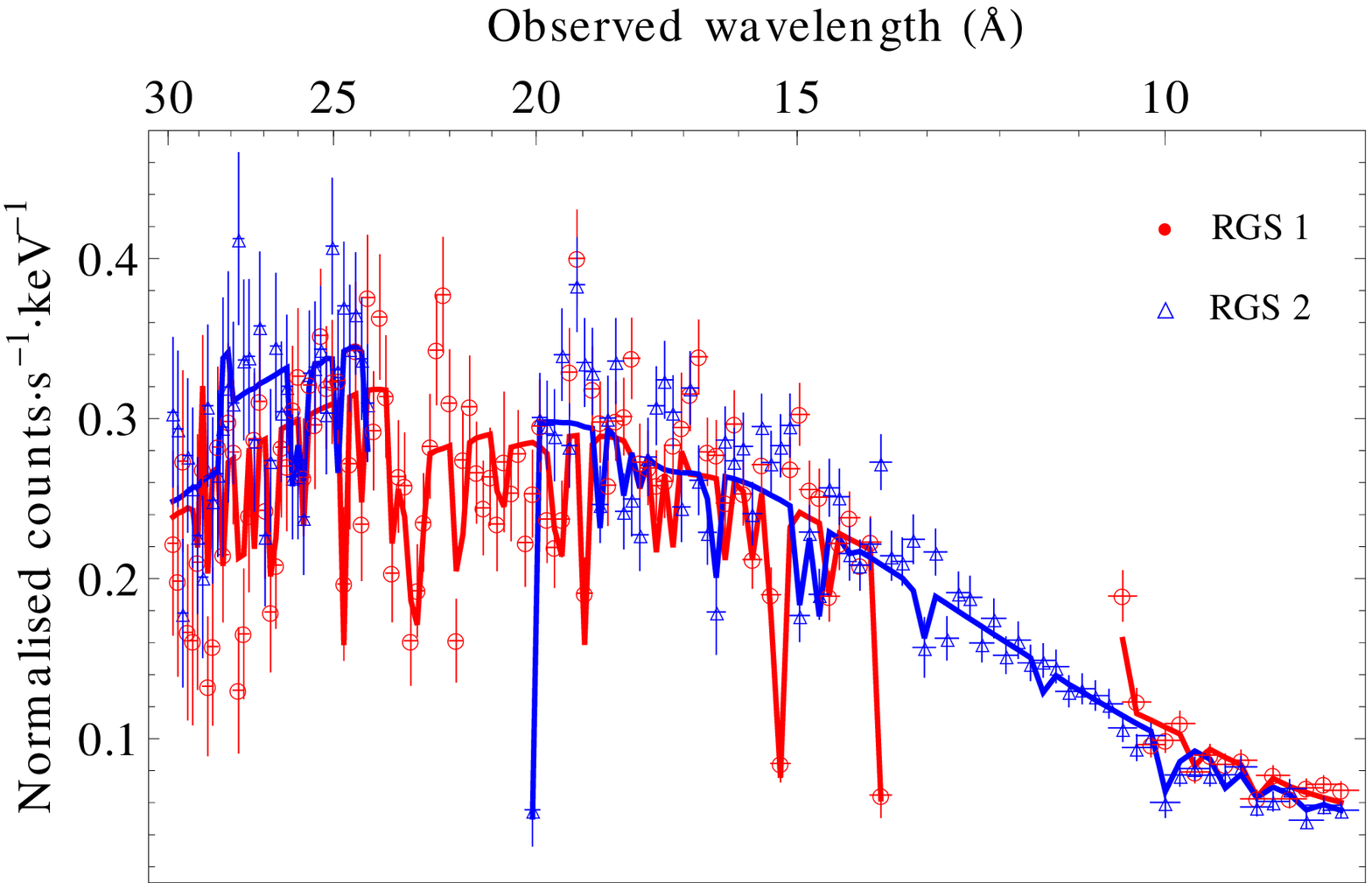}\\[-1.57em]
\hspace*{0.17em}\includegraphics[width=3.483in]{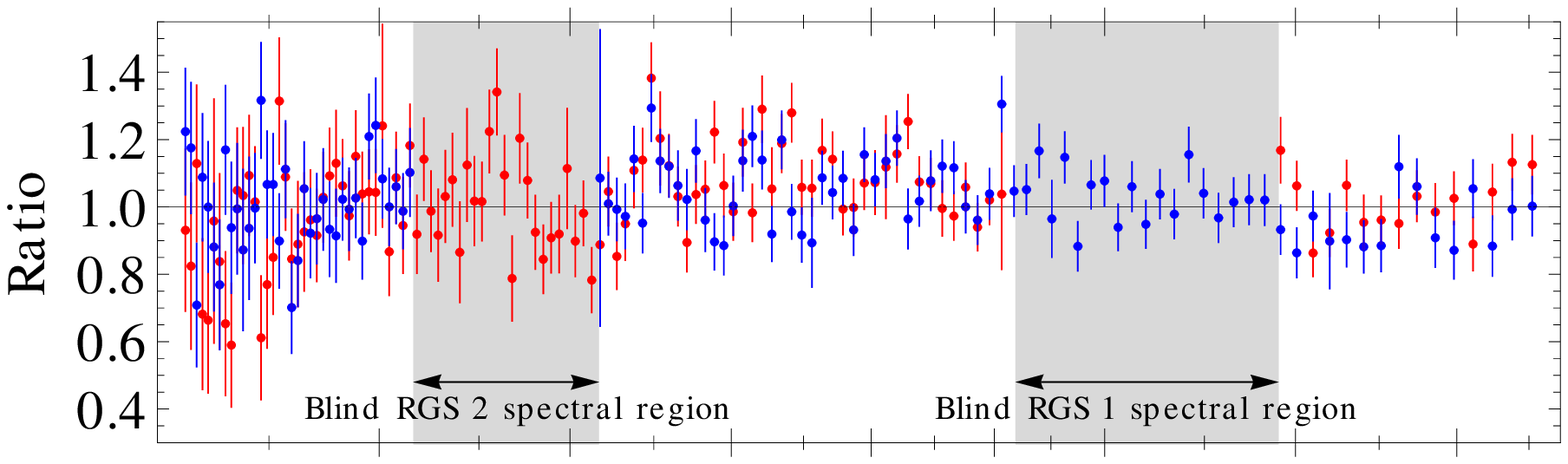}\\[-1.55em]
\includegraphics[width=3.503in]{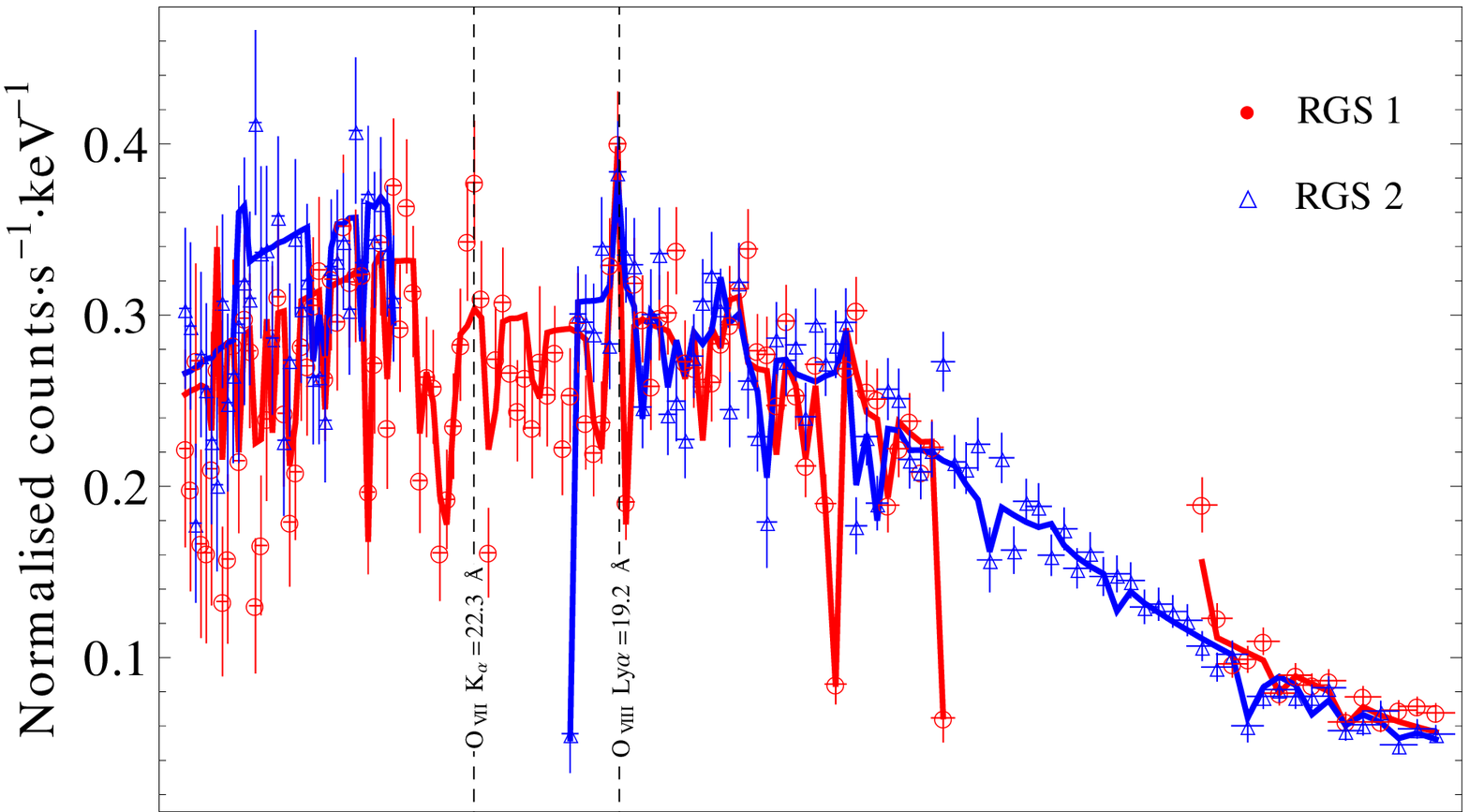}\\[-1.58em]
\hspace*{0.18em}\includegraphics[width=3.483in]{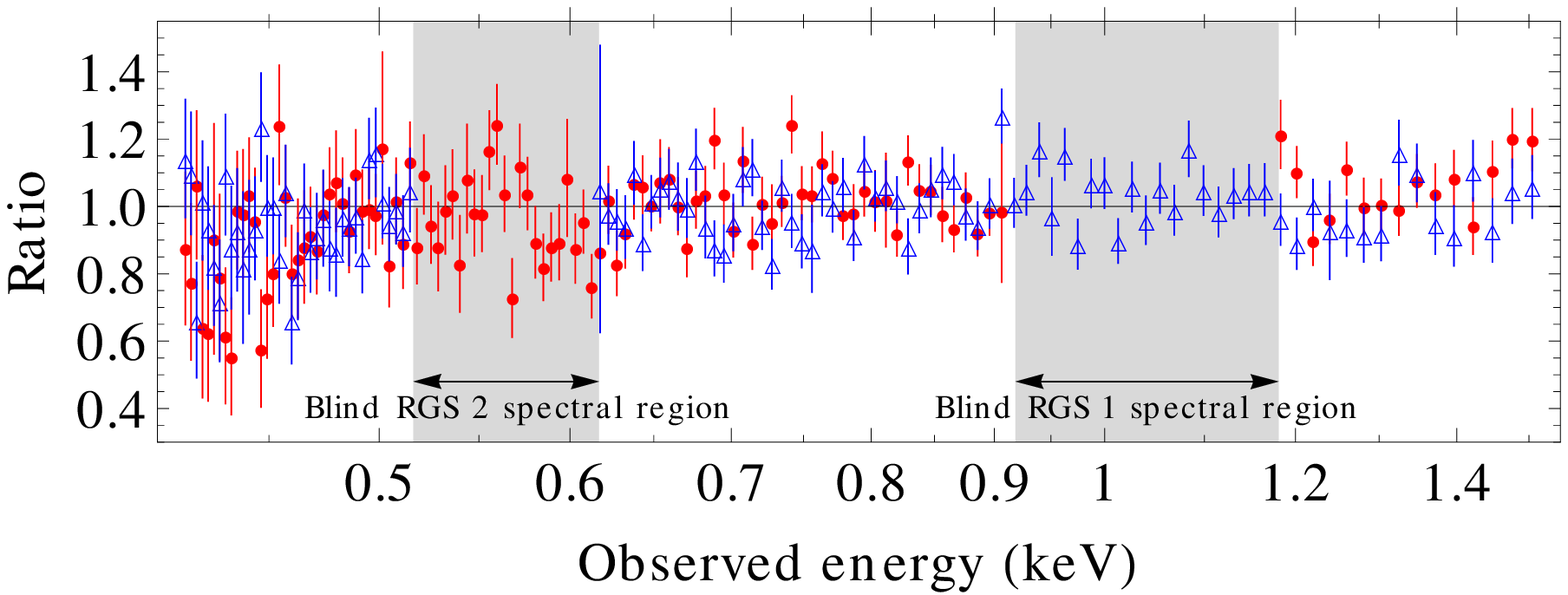}
\caption{RGS spectral fits in the observed 0.4--1.6 keV energy range. The vertical labels indicate the energy in the source's rest-frame. (Top panel) Model A: A redshifted power-law. (Bottom panel) Model B: A thermal component; a redshifted power-law (fixed).}
\label{fig:rgs_SpectralModel}
\end{figure}

Our initial continuum model includes a power-law attenuated by neutral absorption, $N_{\rm H}$ (model A). During the fit, the power-law spectral indices between the two RGSs are linked together in contrast to the power-law normalisations which are left to vary independently between the two RGSs, to account for possible uncertainties in the cross-calibration between the two spectrometers. 
The fit is statistically acceptable with a $\chi^2$ of 298.59 for 269 d.o.f. (0.10 NHP) yielding a best-fitting power-law photon index of $1.78\pm0.03$. However, the ratio plot for both RGSs (Fig.~\ref{fig:rgs_SpectralModel}, top panel) show broad systematic deviations from a constant, as well as two line-like features at around 19 \AA\ (0.653 keV) and 22 \AA\ (0.564 keV, only for RGS\,1), in the observer's frame.\par
In order to cure both the 10--20 \AA\ and 25--30 \AA\ broad excesses in the residuals, and at the same time to model the two line-like emission features, we add to the intrinsic nuclear continuum the {\tt mekal}-model, parametrised by its temperature and normalisation (emission-measure of the gas). Similarly as before both the Hydrogen volume density and the chemical abundance of metals are set to unity and to that of Solar abundance, respectively. During the fit, we link together the temperature of the plasma for the two RGS spectra, while leaving their normalisations to vary independently between them. As for the power-law, we fix the photon index to the best-fitting value derived by Model 5b, 1.86 (Sect.~\ref{ssect:epic_broad_band_Spec}), since the EPIC data trace better the source continuum due to their high sensitivity and energy coverage. This RGS best-fitting model (model B) describes the data significantly better than the simple power-law best-fitting model having a $\chi^2$ of 261.85 for 267 d.o.f. (0.58 NHP) and yielding a best-fitting temperature of $0.39\pm0.07$ keV. The best-fitting model does not leave any significant residuals, associating at the same time the emission lines from the thermal plasma, at 19 and 22 \AA, with a prominent O\,{\sc viii}\,Ly$\alpha$ and O\,{\sc vii}\,K$\alpha$-f, respectively (Fig.~\ref{fig:rgs_SpectralModel}, bottom panel). The best-fitting normalisations of the {\tt mekal} components for the two RGSs are consistent between each other, within their 68 per cent confidence limits, and both of them are also fully consistent with the best-fitting EPIC values. In addition the best-fit temperature is entirely consistent with that of  Model 5b using the EPIC data. From a statistical point of view the addition of the thermal component can be justified by performing an f-test \citep{bevington92} between the best-fitting model A and model B yielding a probability of $1.44\times10^{-6}$ (for an f-statistic value of 14.143) indicating that the addition of the thermal component is valid for 1 per cent significance level.
\par  
The above best-fitting model describes adequately the RGS data of \n7. However, due to the limited RGS spectral range coverage, 0.35--1.8 keV, and to the lack of sufficient sensitivity at the O\,{\sc vii}\,K$\alpha$ and Ne\,{\sc IX}\,K$\alpha$ triplet\footnote{The RGS\,1 is missing the spectral region where Ne\,{\sc IX} lines are usually located (between 11--14 \AA), and the RGS\,2 is missing the region around the O\,{\sc vii} lines (between 20--24 \AA). Both instruments are heavily affected from bad columns. In order to quantify this, we count the number of independent spectral elements ($\delta \lambda= 60$ m\AA) for each RGS affected by bad columns, and we divided it by the total number of independent elements in the analysed bands yielding a fraction of 14 per cent of bad columns for the RGS\,1 and 12 per cent for RGS\,2.} wavelengths, RGSs data alone can neither constrain strongly the existence of plasma with temperatures significantly higher or lower than that of our best-fitting RGS temperature at around 0.3 keV.

\begin{table*}
\begin{minipage}{190mm}
\caption{Best-fit spectral results for the RGS science products in the $[(0.35-0.9) \cup (1.2-1.8)]$ keV.}
\label{tab:RGS_spec_modelFits}
\begin{tabular}{@{}lccccccc}
\hline
Model-parameter & Model A & Model B \\
\hline
1a) {\tt zpowerlw}: Photon index, $\Gamma$ & $1.78\pm0.03$ & $1.86$ \\
1b) {\tt zpowerlw}: Normalisation\textsuperscript{1} & $3.10\pm0.05$  & $2.97^{+0.07}_{-0.05}$ & \\
2a) {\tt mekal}: Temperature, $k_{\rm B}T$ (keV) &  & $0.39\pm0.07$ & \\
2b) {\tt mekal}: H density, $n_{\rm H}$ (cm$^{-3}$) &  & 1 &\\
2c) {\tt mekal}: Metal abundance &  & 1 & \\
2d) {\tt mekal}: Normalisation\textsuperscript{2} &  & $6.21^{+2.21}_{-2.07}$\textsuperscript{\textasteriskcentered} &\\
\hline
$\chi^2/$d.o.f. & 289.59$/$269 & 261.85$/$267 \\
\hline
\end{tabular}
\medskip\\
Values without uncertainties correspond to fixed fitting parameters.\\
The best-fitting normalisations correspond to the average of the two RGSs.\\
\textsuperscript{\textasteriskcentered} Individually the normalisations for RGS\,1 and RGS\,2 are $7.47\pm2.12$ and $4.93^{+2.30}_{-1.84}$ respectively (in the corresponding units).\\
\textsuperscript{1} In units of 10$^{-3}$ phot. keV$^{-1}$cm$^{-2}$ s$^{-1}$ at 1 keV.\\
\textsuperscript{2} In units of $\int n_{\rm e}\,n_{\rm H} dV$, ($10^{62}$ cm$^{-3}$).
\end{minipage}
\end{table*}

\section{SUMMARY AND DISCUSSION}
\label{sect:discus}

We have analysed the longest \textit{XMM-Newton} observation of \n7, obtained during November 2009. We have performed hardness ratio analysis, on the EPIC light curves in different energy bands, yielding no evidence of X-ray spectral variability within the given observation. This enabled us to perform a detailed X-ray spectral analysis on both the EPIC and the RGS data products using the total X-ray spectra, in the 0.3--10 keV and 0.35--1.8 keV energy bands, respectively.\par
The main results from our X-ray spectral analysis can be summarised as follows.
\begin{itemize}
\item The EPIC X-ray spectrum:
\begin{itemize}
 \item above 3 keV exhibits three narrow emission iron lines at $6.41\pm0.01$, $6.71\pm0.02$ and  $6.96^{+0.03}_{-0.05}$ being consistent with the neutral \fa (EW: $126^{+32}_{-27}$ eV), the Fe {\sc xxv} resonance (EW: $58^{+31}_{-20}$ eV) and the 1s--2p doublet Fe {\sc xxvi} K$\alpha$ (EW: $23^{+29}_{-23}$ eV).
 \item in the full band (0.3--10 keV) can be best-fitted by an underlying power-law model with $\Gamma=1.86\pm0.02$, and two thermal models, ({\tt mekal}), of temperatures equal to $k_{\rm B}T_{1}=0.36\pm0.04$ keV and $k_{\rm B}T_{2}= 8.84^{+1.78}_{-1.11}$ keV (with the highest temperature component effectively accounting for the ionised iron emission lines)
\end{itemize}
\item The RGS X-ray spectrum:
\begin{itemize}
\item The results from the RGS X-ray spectrum analysis in the 0.35--1.8 keV band are fully consistent with the best-fitting model of the broad band EPIC spectrum. In particular, the RGS spectrum can be well fitted by a power-law component and a {\tt mekal} component of temperature $k_{\rm B}T=0.39\pm0.07$ keV.
\end{itemize}
\end{itemize}

The best-fitting photon spectral index that we get from the EPIC data together with the flux estimates (Model 5b, Sect.\ref{ssect:epic_broad_band_Spec}), are fully consistent, within the 68 per cent confidence limits, with the linear anticorrelation relation derived by \citet{emmanoulopoulos12}. The latter is deduced using all the archival RXTE data, showing unambiguously that \n7 behaves similarly to Galactic `hard state' sources exhibiting a `harder when brighter' X-ray behaviour.\par
The EW value of the neutral \fa is significantly larger than the one derived from earlier \textit{XMM-Newton} observations \citep[around $82\pm10$ eV;][]{bianchi03,starling05} and \textit{Suzaku} observations \citep[around $83\pm10$ eV;][]{lobban10} (note that the error bars indicate the 90 per cent confidence limits of the corresponding EWs). However, it is absolutely in accordance with the one derived from the \textit{Chandra}-HETG \citep[$120^{+40}_{-30}$ eV;][]{bianchi08}. In the case of X-ray reflection from Compton-thick material in a plane-like configuration, the EW of the narrow iron emission line should be tightly connected with the reflection strength $R$. In fact, one considers a linear correlation between the two parameters, being equal to unity when the \fa line's EW is 120 eV \citep[e.g.][]{matt91,bianchi09}, then the EW value of the neutral \fa line that we measure implies a reflection strength of $R\simeq1$, corresponding to an average solid angle, subtended by the Compton-thick reflector of $2\pi$ sr. This is not consistent with our results disfavouring the Compton-thick reflector origin of neutral \fa.\par
This result favours scenarios in which the neutral \fa line originates from a Compton-thin material, such as the BLR \citep[e.g.][]{bianchi08}. Following the procedure of \citet{yaqoob01}, assuming a spherically symmetric cloud distribution and neutral iron, the expected EW for the BLR in \n7 can be written (in a similar fashion to \citet{bianchi08}) as 
\begin{eqnarray}
{\rm EW}_{{\rm Fe\;K}\alpha }=31.7\left(\frac{f_{\rm c}}{0.35}\right)\left(\frac{N_{\rm h}}{10^{23}\rm{cm}^{-2}}\right)\,\rm{eV}
\end{eqnarray}
where $f_{\rm c}$ is the covering factor and $N_{\rm h}$ the hydrogen column density of each cloud. For this relation we assume an iron abundance relative to hydrogen of $A_{\rm Fe}=4.68\times10^{-5}$ \citep{anders89} and the fluorescence yield is set to $\omega_{\rm k}=0.342$ for neutral iron \citep{bambynek72}. Assuming $f_{\rm c}=0.35$ \citep{goad98} a column density of $N_{\rm h}=4\times10^{23}$ cm$^{-2}$ is required to reproduce the \fa's EW. These values of  $f_{\rm c}$ and $N_{\rm h}$ are commonly expected from the BLR \citep{netzer90}, making the latter indeed a possible candidate for the origin of the neutral \fa emission line in \n7.\par
The X-ray spectrum of \n7 is known to exhibit highly ionised iron emission lines  between 6.5--7 keV \citep{bianchi02,bianchi05,starling05,bianchi08,lobban10}. From our EPIC data, we find that the dominant component of the He-like Fe {\sc xxv} triplet is the resonant one. This is in agreement with both older \textit{XMM-Newton} data \citep{starling05} and \textit{Chandra}-HETG data \citep{bianchi08} but not with the \textit{Suzaku} data \citep{lobban10}, the latter associating it with the forbidden transition. The resonant nature of Fe {\sc xxv} line indicates that it is produced within a gas which is collisionally excited \citep{porquet00,bautista99}.\par
Finally, both the EPIC and the RGS data suggest the presence of a multi-temperature hot gas in the nuclear region of \n7. Our broad band EPIC results show that the source's `soft excess' component, at energies below 2 keV, can be well fitted by one thermal component with best-fitting temperature of around 0.36 keV, something which is in accordance with the RGS data. Previous results of \n7 from \citet{lobban10} consider two thermal emission components to best-fit the low-energy part of the X-ray spectrum at 0.24 and 0.86 keV. \citet{starling05} found evidence for a single low energy thermal component at 0.18 keV and by fitting a second one, at 0.56 keV, the fit was not significantly improved. Both the pn and MOS spectra from the long {\it XMM-Newton} observation that we analyse in this work are not consistent with the presence of a second thermal component. Interestingly, our best-fit temperature of the single low-energy thermal component is intermediate between that reported by \citet{lobban10}. This could perhaps explain the difference between ours and the previously reported results. Finally, the thermal component with the higher temperature, 8.84 keV, can effectively account for the previously mentioned ionised iron emission lines, that we detect between 6.5--7 keV, being indicative for a gas in collisional ionisation equilibrium.\par
The {\it Chandra} X-ray image of the source \citep{bianchi08} shows only an unresolved nuclear source (even at low energies) within a few arcseconds around the galactic nucleus. Consequently, if a multi-temperature gas does exist in this source, it has to be confined in a small region, close to the AGN. Perhaps then, intense starburst superwind activity in the nuclear region of this source could account of this multi-temperature medium, which may even be outflowing at velocities of the order of 1000 km s$^{-1}$, as suggested by the iron line's blueshifts measured by \citet{bianchi08}. In the framework of superwinds, we expect the hard X-ray emission from the galaxy driving the superwind coming from the hot and tenuous supernova/wind-heated gas inside the starburst (i.e.\ inside the sonic radius of the wind) \citep{heckman93,strickland01}. In this environment, characteristic temperatures between 6--9 keV can arise yielding in a self-consistent way the observed ionised iron species.

\section*{Acknowledgments}
DE and IMM acknowledge the Science and Technology Facilities Council (STFC) for support under grant ST/G003084/1. IP and FN acknowledge support by the EU FP7-REGPOT 206469 grant. We would like to thank A.~P.~Lobban for useful discussions on his manuscript. This research has made use of NASA's Astrophysics Data System Bibliographic Services. Finally, we are grateful to our referee, M\'{o}nica Cardaci, for her useful comments that helped improved the quality of the manuscript substantially.

\bibliographystyle{mnauthors}

\bsp
\label{lastpage}
\end{document}